\documentclass[11pt]{article} %twocolumn
\usepackage{jcappub,natbib,float,caption,comment,bm}
\usepackage{color}
\input{colordvi.tex}

\newcommand{\calA}{{\cal A}}

\newcommand{\calG}{{\cal G}}

\newcommand{\calL}{{\cal L}}
\newcommand{\calN}{{\cal N}}
\newcommand{\calO}{{\cal O}}
\newcommand{\calP}{{\cal P}}
\newcommand{\calQ}{{\cal Q}}

\newcommand{\calV}{{\cal V}}
\newcommand{\dd}{{\rm d}}
\newcommand{\pa}{\partial}
\newcommand{\vphi}{\varphi}
\newcommand{\vep}{\varepsilon}
\newcommand{\he}{\spadesuit}

\newcommand{\ma}[1]{{\mathrm{#1}}}
\newcommand{\vect}[1]{\mbox{\boldmath${#1}$}}

\title{Anisotropic inflation reexamined: upper bound on broken
rotational invariance during inflation}
\author[a,b]{Atsushi Naruko,}
\author[c,d]{Eiichiro Komatsu,}
\author[a]{Masahide Yamaguchi}
\affiliation[a]{Department of Physics, Tokyo Institute of Technology,
 2-12-1 Ookayama, Meguro-ku, Tokyo 152-8551, Japan}
\affiliation[b]{APC (CNRS-Universit\'e Paris 7),
 10 rue Alice Domon et L\'eonie Duquet, 75205 Paris Cedex 13, France}
\affiliation[c]{Max-Planck-Institut f\"ur Astrophysik,
 Karl-Schwarzschild-Str. 1, 85741 Garching, Germany}
\affiliation[d]{Kavli Institute for the Physics and Mathematics of
 the Universe (WPI), Todai Institutes for Advanced Study,
 The University of Tokyo, 5-1-5 Kashiwanoha, Kashiwa, Chiba 277-8583, Japan}
\emailAdd{naruko@th.phys.titech.ac.jp}

\abstract{
The presence of a light vector field coupled to a scalar field during
inflation makes a distinct prediction: the observed correlation
functions of the cosmic microwave background (CMB) become statistically
anisotropic. We study the implications of the current bound on
statistical anisotropy derived from the Planck 2013 CMB temperature data
for such a model. The previous calculations based on the attractor
solution indicate that the magnitude of anisotropy in the power spectrum
is proportional to $N^2$, where $N$ is the number of $e$-folds of
inflation counted from the end of inflation. In this paper, we show that
the attractor solution is not necessarily compatible with the current
bound, and derive new predictions using another branch of anisotropic
inflation.  In addition, we improve upon the calculation of the mode
function of perturbations by including the leading-order slow-roll
corrections. We find that the anisotropy is roughly proportional to
$[2(\vep_H+4\eta_H)/3-4(c-1)]^{-2}$, where $\vep_H$ and $\eta_H$ are the
usual slow-roll parameters and $c$ is the parameter in the model,
regardless of the form of potential of an inflaton field. The bound from
Planck implies that breaking of rotational invariance during inflation
(characterized by the background homogeneous shear divided by the Hubble
rate) is limited to be less than ${\cal O}(10^{-9})$. This bound is many
orders of magnitude smaller than the amplitude of breaking of time
translation invariance, which is observed to be ${\cal O}(10^{-2})$.}
\begin{document}

\maketitle
%%%%%%%%%%%%%%%%%%%%%%%%%%%%%%%%%%%%%%%%%%%%%%%%%%%%%%%%%%%%%%%%%%%%%%%%%%
%%%%%%%%%%%%%%%%%%%%%%%%%%%%%%%%%%%%%%%%%%%%%%%%%%%%%%%%%%%%%%%%%%%%%%%%%%
\section{Introduction}
Invariance of the probability distribution of primordial curvature
perturbations under spatial rotation and translation is the fundamental
prediction of the standard single scalar-field inflation models
\cite{Mukhanov:1990me}. On the other hand, the probability distribution
is only approximately invariant under spatial dilation; thus, the
two-point correlation function of the primordial curvature perturbation
is approximately, but not exactly, scale invariant
\cite{Mukhanov:1981xt}. Deviation from the exact scale invariance has
been detected conclusively with more than 5 standard deviations
\cite{Hinshaw:2012aka,Ade:2013zuv} which, along with stringent limits on
deviation from Gaussian statistics \cite{Bennett:2012zja,Ade:2013ydc},
supports the idea that the observed structures in the universe originate
from quantum fluctuations generated during inflation.

These standard predictions depend little on details of the single-field
inflation models, as the statistical properties of the probability
distribution are determined primarily by symmetry of spacetime; namely,
spacetime during inflation is nearly de Sitter with the Hubble expansion rate
slowly varying with time, i.e., $|\dot H/H^2|={\cal O}(10^{-2})$.
Nearly, but not exactly, scale-invariant correlation function is the
consequence of this. Then, the natural question is, ``what if spacetime during
inflation is slightly anisotropic, like a Bianchi type?'' How much do we
know about violation of rotational symmetry during inflation? Can we
place a bound on it?

{\it Anisotropic inflation} is a class of multi-field inflation
models. They contain a vector field violating rotational symmetry (see
\cite{Dimastrogiovanni:2010sm,Maleknejad:2012fw,Soda:2012zm} for
reviews).  As a result, $N$-point correlation functions of the curvature
perturbation become statistically anisotropic
\cite{Ackerman:2007nb,Watanabe:2010fh,Barnaby:2012tk,Bartolo:2012sd,Shiraishi:2013vja,Shiraishi:2013oqa,Emami:2013bk,Abolhasani:2013zya,Emami:2014tpa}.\footnote{Anisotropic
phase prior to inflation \cite{Dey:2011mj,Dey:2012qp,Dey:2013tfa} and
the so-called solid inflation
\cite{Endlich:2012pz,Bartolo:2013msa,Bartolo:2014xfa,Akhshik:2014gja} can
also generate anisotropies in the correlation functions.}  In this
paper, we shall focus on anisotropy in the two-point correlation
function sourced by an $F^2$ term in the action. While this model has
been studied extensively in the literature
\cite{Watanabe:2009ct,Watanabe:2010fh,Barnaby:2012tk,Bartolo:2012sd,Dimopoulos:2010xq},
 we show that there are subtleties missing in the literature which results
in a revised prediction of this particular model. We also obtain an
upper bound on violation of rotational symmetry during inflation in this
model, using the latest bound from the Planck 2013 temperature anisotropy data
\cite{Kim:2013gka}.

This paper is organized as follows. In section~\ref{sec:newbranch}, we
show that the attractor solution of the $F^2$ model, on which the
previous calculations
\cite{Watanabe:2010fh,Barnaby:2012tk,Bartolo:2012sd} are based, is
incompatible with the current bound on statistical anisotropy
\cite{Kim:2013gka} except for the special case with fine-tuning,
and then discuss a new branch of anisotropic inflation, particularly
paying attention to the background solution.  In
section~\ref{sec:perturbations}, cosmological perturbations on a new
background solution are discussed and their second-order actions are
obtained. In section~\ref{sec:anisotropy}, we calculate the statistical
anisotropy. In section~\ref{sec:rotationalinv}, we discuss implications
of our results for the bound on breaking of rotational invariance during
inflation. We conclude in section~\ref{sec:conclusion}. We give detailed
derivation of the perturbed action in the appendix. The reduced
Planck scale $M_{\rm pl} = 1 / \sqrt{8\pi G}$ is set to be unity unless
otherwise specified.
%%%%%%%%%%%%%%%%%%%%%%%%%%%%%%%%%%%%%%%%%%%%%%%%%%%%%%%%%%%%%%%%%%%%%%%%%%
%%%%%%%%%%%%%%%%%%%%%%%%%%%%%%%%%%%%%%%%%%%%%%%%%%%%%%%%%%%%%%%%%%%%%%%%%%
\section{New branch in anisotropic inflation}
\label{sec:newbranch}
%%%%%%%%%%%%%%%%%%%%%%%%%%%%%%%%%%%%%%%%%%%%%%%%%%%%%%%%%%%%%%%%%%%%%%%%%%
\subsection{Motivation}
Watanabe, Kanno and Soda \cite{Watanabe:2009ct} found the first
working model (e.g., free from a ghost) of inflation with a vector field
that can produce persistent anisotropy in the background spacetime.
This model provides a counter-example to the ``cosmic no-hair
conjecture,'' which states that the spacetime rapidly approaches 
quasi de Sitter spacetime during inflation.
The same authors found a solution of anisotropic spacetime which is
an attractor (i.e., the solution that is independent of initial
conditions of scalar (inflaton) and vector fields). The attractor
solution yields a time-independent ratio of the energy densities of the vector
and scalar fields.

However, as we shall show in this paper, the conditions in which the
attractor solution is valid are inconsistent with the current
observational bound on $g_*$ \cite{Kim:2013gka} unless the 
 initial conditions, e.g. those for the matter fields, are fine-tuned so that
 the attractor solution is realized from the beginning of inflation. 
The goal of this paper is to find another branch of
solutions that is compatible with the observational bound, and derive
new predictions of this model for statistical properties of
perturbations.

The action of the model is given by \cite{Watanabe:2009ct}
 \begin{align}
 S = \int \dd x^4 \sqrt{- g} \left[ \frac{1}{2} R
 - \frac{1}{2} g^{\mu \nu} \pa_\mu \phi \pa_\nu \phi - U (\phi)
 - \frac{1}{4} f^2 (\phi) F_{\mu \nu} F^{\mu \nu} \right]  \,,
\label{action}
 \end{align}
where $R$ is the Ricci scalar, $U (\phi)$ is the potential of an
inflaton field, $F_{\mu \nu} \equiv \pa_\mu A_\nu - \pa_\nu A_\mu$ is
the field strength tensor of the vector field $A_\mu$, and $f (\phi)$ is
the coupling function between $\phi$ and $A_\mu$. 
In this subsection, we shall follow ref.~\cite{Watanabe:2009ct} and
assume the following particular forms of the potential and the coupling
function:
 \begin{align}
 U (\phi) = \frac{1}{2} m^2 \phi^2 \,, \qquad
 f (\phi) = e^{\frac{1}{2} c \phi^2} \,,
\label{pot-coup}
 \end{align}
where $m$ is an inflaton mass and $c$ is a constant. Note that this
form of the potential is used in this subsection only. The analysis in
the later sections is completely general, and is applicable to any forms
of the potential. The coupling function shall be determined by the form
of the potential. Using the slow-roll approximation, $3 H^2 \approx U
(\phi)$ and $\ddot{\phi} \ll H \dot{\phi}$ where $H$ is the Hubble rate,
the scalar field equation can be simplified as
\begin{align}
 \frac{\dd}{\dd \alpha} \phi
 \approx - \frac{U_\phi}{U} (1 - c I) \,,
 \qquad
 \dd \alpha \equiv H \dd t \,. 
\label{eq:scalar-jiro}
\end{align}
Because of the presence of the coupling $f (\phi)$, the vector field
affects the equation of motion of the scalar field through a new
function $I$ defined by
 \begin{align}
 I \equiv
 2 \left( \frac{1}{2} \frac{U_\phi^2}{U^2} \right)^{- 1}
 \frac{\rho^{\vec{A}}}{U}  
 = \frac{I_\he}{e^{c (\phi^2 - \phi_\he^2) + 4 (\alpha - \alpha_\he)}} \,.
\label{eq:I}
 \end{align}
Here, $\rho^{\vec{A}}$ is the energy density of the vector field, and
the subscript $\he$ denotes a certain epoch during inflation, $t =
t_\he$.  

As shown by ref.~\cite{Watanabe:2010fh}, $I$ determines the amplitude of
statistical anisotropy in the power spectrum of the curvature
perturbation. When we write the power spectrum as \cite{Ackerman:2007nb}
\begin{align}
 P ({\bm k})
 = P (k) \Bigl[ 1 + g_* (\hat{{\bm k}} \cdot \hat{{\bm v}})^2 \Bigr]\,,
\end{align}
where $\hat{\bm v}$ is some preferred direction in space, the amplitude
of statistical anisotropy, $g_*$, is related to $I$ via
\begin{align}
 g_* = - 24 I N_k^2 \,.
\label{g:Jiro}
\end{align}
Here, $N_k$ is the number of $e$-folds of inflation counted from the end
of inflation to the epoch at which a perturbation with a given
wavenumber, $k = |{\bm k}|$, left the horizon.

Solving equation~(\ref{eq:scalar-jiro}), the general solution of $I$
 up to slow-roll corrections is given by
\begin{align}
 I = \frac{c - 1}{c^2} \frac{1}{1 + \left( \frac{c - 1}{c^2}
 \frac{1}{I_\he} - 1 \right) e^{- 4 (c - 1) (\alpha - \alpha_\he)}} \,.
\label{sol:Jiro}
\end{align}
If $c - 1 \geq \calO (1)$ and $I_\he$ is not extremely small, the
second term in the denominator can be neglected, and $I$ is simply
determined by the model parameter $c$ as
 \begin{align}
 I \to \frac{c - 1}{c^2}\,.
 \end{align}
This is the attractor solution found by
ref.~\cite{Watanabe:2009ct}. Using equation~(\ref{g:Jiro}), we can
compare the prediction of the attractor solution for $g_*$ directly with
the observation. The current bound on $g_*$ is given by
\cite{Kim:2013gka}
$g_* = 0.002 \pm 0.016$ (68\%~CL), which yields the constraint on $c$ as
 \begin{align}
 I = \frac{c - 1}{c^2} \approx c - 1 \lesssim
 10^{-7} \times \left( \frac{|g_*|}{10^{-2}} \right)
 \left( \frac{N_k}{60} \right)^{- 2}\,.
\label{eq:boundonI}
 \end{align}
We thus find that $c$ must be extremely close to unity. This result
contradicts with the assumption made to obtain the attractor solution in
the first place, i.e. $c - 1 \geq \calO (1)$. Therefore, we must
conclude that the attractor solution is inconsistent with the
observation.\footnote{While we used the formula for $g_*$ derived in the
previous work ref.~\cite{Watanabe:2010fh} to reach this conclusion,
their formula needs a correction as we shall show in
section~\ref{sec:anisotropy}. We have checked that the corrected formula
for $g_*$ still gives the same conclusion that the attractor solution
for the background is not consistent with the observational bound on
$g_*$ except for the special case with fine-tuning.} 

In passing, one might wonder from equation~(\ref{sol:Jiro}).
 if the attractor mechanism works well as long as inflation lasts
 long enough in the case $c > 1$.
 Unfortunately, if inflation occurs at very high energy scale and
lasts long enough, one cannot neglect quantum-mechanically generated
vector field, which can easily overcome the background vector field as
will be discussed in subsection~\ref{subsec:QGV}. Thus, given extremely
small $c-1$, inflation cannot last long enough to realize the attractor
solution for a wide range of initial conditions, as long as the
perturbative treatment is justified. Therefore, we consider a new
solution instead of the attractor solution in this paper.

%%%%%%%%%%%%%%%%%%%%%%%%%%%%%%%%%%%%%%%%%%%%%%%%%%%%%%%%%%%%%%%%%%%%%%%%%%
\subsection{New solution for the anisotropic background}

In anisotropic inflation, the background spacetime is homogeneous but
{\it anisotropic} because of the presence of a vector field. To describe
the solution, we take the line element of Bianchi type I spacetime with
the following form:
\begin{align}
 \dd s^2 = - \calN^2 (t) \dd t^2
 + e^{2 \alpha (t)} \Bigl[ e^{- 4 \beta (t)} \, \dd x^2
 + e^{2 \beta (t)} \, \delta_{a b} \, \dd y^a \dd y^b \Bigr] \,,
\end{align}
where $e^\alpha$ is the standard isotropic scale factor in an
isotropic background, and $e^\beta$ describes anisotropy in the
expansion. The indices, $a$ and $b$, take on $2$ and $3$ (i.e., $y$ and
$z$ axes).

In the model of ref.~\cite{Watanabe:2009ct}, there exists a homogeneous
vector field at the onset of inflation. Without loss of generality, we
shall take the direction of the initial vector field to be the $x$-axis,
i.e., $A_\mu = (0, u (t), 0, 0)$. One can question the origin of such a
homogeneous vector field at the onset of inflation. If inflation starts,
quantum fluctuations in $A_\mu$ inside the horizon would be stretched to
outside the horizon, and provide a classical background $u(t)$. However,
it is also natural to expect that there were fluctuating vector fields
before inflation, whose gradient energy density would prevent inflation
from starting. In fact, the situation is not different from the standard
inflation only with a scalar field. To start inflation, one must
have a scalar field which is homogeneous over a few Hubble radius at the
onset of inflation, but we do not know how to arrange such a homogeneous
field at the beginning without fine-tuning or anthropic
argument. Therefore, we shall postpone this challenging question
regarding the origin of homogeneous vector and scalar fields at the
onset of inflation, and proceed.

The dynamics of $u (t)$ is governed by the following equation of motion:
\begin{align}
 0 = \frac{1}{\sqrt{- g}} \pa_\nu \Bigl( \sqrt{- g} f^2 F^{\mu \nu} \Bigr)
 = \frac{1}{e^{3 \alpha}} \pa_t \Bigl( f^2 e^{\alpha + 4 \beta} \, \dot{u} \Bigr)
 \delta^\mu{}_x \,.
\end{align}
Integrating this equation yields
\begin{align}
 \dot{u} = \frac{C_A}{f^2 e^{\alpha + 4 \beta}} \,,
\end{align}
with $C_A$ being an integration constant related to the initial
condition of the vector field. The energy density, $\rho^{\vec{A}}$,
isotropic pressure, $P^{\vec{A}}$, and anisotropic stress,
$(\pi^{\vec{A}})^{i}_{j}$, of the vector field are given by
 \begin{align}
 \rho^{\vec{A}} = \frac{1}{2} \calV \,, \qquad
 P^{\vec{A}} = \frac{1}{6} \calV \,, \qquad
 \bigl( \pi^{\vec{A}} \bigr)^x{}_x = - \frac{2}{3} \calV \,, \qquad
 \bigl( \pi^{\vec{A}} \bigr)^a{}_b = \frac{1}{3} \calV \, \delta^a{}_b \,,
 \end{align}
where
 \begin{align}
 \calV \equiv \frac{f^2 \dot{u}^2}{e^{2 (\alpha - 2 \beta)}}
 = \frac{C_A^2}{f^2 e^{4 (\alpha + \beta)}} \,.
 \end{align}
Using this function ${\cal V}$, the background Einstein equations read
\begin{subequations}
\begin{align}
 3 H^2 - 3 \dot{\beta}^2
 &= \frac{1}{2} \dot{\phi}^2 + U (\phi) + \frac{1}{2} \calV \,,
\label{eq:hami} \\
 \dot{H} + 3 H^2
 &= U (\phi) + \frac{1}{6} \calV \,, \\
 \ddot{\beta} + 3 H \dot{\beta}
 &= \frac{1}{3} \calV \,,
\label{eq:beta} 
\end{align}  
\end{subequations}
and the equation of motion of $\phi$ reads
 \begin{align}
 \ddot{\phi} + 3 H \dot{\phi} + U_\phi
 &= \frac{f_\phi}{f} \calV \,.
\label{eq:scalar}
 \end{align}

We find the background solution in the leading order of the slow-roll
expansion. First, to realize anisotropic inflation, let us relate the
form of the coupling function, $f (\phi)$, to an arbitrary inflaton
potential, $U(\phi)$, as in ref.~\cite{Watanabe:2009ct}\,:
\begin{align}
 f (\phi) = f_\he \, \exp \left[ 2c \int_{\phi_\he}^\phi \dd \phi'
 \frac{U}{U_\phi} \right] \,,
\label{def:fphi}
\end{align}
where $c$ is the same constant introduced in equation
(\ref{pot-coup}). Then the right hand side of the equation of motion
of $\phi$ (equation~(\ref{eq:scalar})) becomes
 \begin{align}
 \ddot{\phi} + 3 H \dot{\phi} + U_\phi
 &= 2 c \frac{U}{U_\phi}\calV \,.
\label{eq:scalar2}
 \end{align}
The function determining the statistical anisotropy, $I$
(equation~(\ref{eq:I})), now reads
\begin{align}
 I \equiv 2 \left( \frac{1}{2} \frac{U_\phi^2}{U^2} \right)^{- 1}  
 \frac{\rho^{\vec{A}}}{U}
 = \left( \frac{1}{2} \frac{U_\phi^2}{U^2} \right)^{- 1} \frac{\calV}{U}
 \,.  
\end{align}
Physically, this function $I$ is the ratio of the energy densities
of the vector and scalar fields, divided by the slow-roll parameter.

So far, the basic equations are general, except for the fact that the
form of $f (\phi)$ is fixed by the potential as in
equation~(\ref{def:fphi}). We can, in principle, consider a case with
arbitrarily strong anisotropy by solving the basic equations without any
approximations. However, according to the previous formula of
ref.~\cite{Watanabe:2010fh}, the current bound on $g_*$ implies that $I$
is smaller than of order $10^{-7}$ (equation~(\ref{eq:boundonI})).
While this result was obtained using the attractor solution that was not
compatible with $I\ll 1$ except for the special case with
fine-tuning, we shall still assume that $I$ is small, and solve the
basic equations by expanding them in terms of $I$. Physically, we assume
the ratio of the energy densities of vector and scalar fields to be
smaller than the slow-roll parameter. For a small $I$, we find a general
solution for $I$ regardless of the form of $U(\phi)$:
\begin{align}
 I = I_\he \, e^{4 (c - 1) (\alpha - \alpha_\he)} \,,
 \qquad
 \left. I_\he \equiv 2 \frac{U}{U_\phi^2} \frac{C_A^2}{f^2 \,e^{4 \alpha}}
 \right|_{t = t_\he} \,.
 \label{sol:I}
\end{align}
where we have used the standard slow-roll conditions,
$3 H^2 \approx U (\phi)$ and $\ddot{\phi} \ll H \dot{\phi}$, and ignored
the slow-roll terms. We shall include the leading-order slow-roll
terms when studying perturbations in the next section.

Under the slow-roll approximation, the scalar field
equation~(\ref{eq:scalar}) reads
 \begin{align}
 3 H \dot{\phi}+ U_\phi = c \, U_\phi I \quad \to \quad
 \frac{\dd \phi}{\dd \alpha} = - \frac{U_\phi}{U} (1 - c I)
 \approx - \frac{U_\phi}{U} \,.
\end{align}
Using this result in equation~(\ref{def:fphi}) and integrating, we obtain
\begin{align}
 f (\phi) = f_\he \, \exp \left[ - 2c \int_{\phi_\he}^\phi \dd \phi'
 \frac{\dd \alpha}{\dd \phi'} \right]
 = f_\he \, e^{- 2 c (\alpha - \alpha_\he)} \,.  
\end{align}
Using this result in equation~(\ref{sol:I}), we obtain the explicit
solution for $I$ with arbitrary $U(\phi)$. While the previous attractor
solution, $I_{\rm attractor}=(c-1)/c^2$, was independent of time, the
new solution depends on time. The parameter $c$ now determines the rate
of evolution of $I$, rather than the value of $I$ itself.

%%%%%%%%%%%%%%%%%%%%%%%%%%%%%%%%%%%%%%%%%%%%%%%%%%%%%%%%%%%%%%%%%%%%%%%%%%
%%%%%%%%%%%%%%%%%%%%%%%%%%%%%%%%%%%%%%%%%%%%%%%%%%%%%%%%%%%%%%%%%%%%%%%%%%
\section{Perturbations}
\label{sec:perturbations}

\subsection{Scalar and vector modes}
In this section, we study cosmological perturbations to the homogeneous
but anisotropic background solution obtained in the last
section. Different from the Friedmann-Lema\^itre-Robertson-Walker (FLRW)
case, where the spacetime is homogeneous {\it and} isotropic, our
background solution has less symmetry. Our solution has only $2$D
rotational symmetry as opposed to $3$D in the standard FLRW
set-up. The perturbations can be classified based on $2$D symmetry in
the $y$-$z$ plane; thus, they have $2$D-scalar and $2$D-vector types.

An arbitrary $2$D-vector, $v^a$, can be uniquely decomposed into one
scalar and one vector modes. We can extract the scalar mode from the
vector by taking divergence, and then obtain the vector mode by
subtracting the scalar part from the original vector. In the same way,
a $2$D tensor, $t_{a b}$, can be decomposed into two scalar and one vector
modes. Therefore, all of the perturbations are classified into scalar
and vector modes as follows. From the matter fields (scalar and vector
fields) we have
 \begin{align}
 \ma{scalar} : \delta \phi \,, \delta A_0 \,, \delta A_x \,,
 1 \, \ma{of} \, \delta A_a \,,  \qquad
 \ma{vector} : 1 \, \ma{of} \, \delta A_a \,,  
 \end{align}
 and from the metric we have
 \begin{align}
 \ma{scalar} : \delta g_{0 0} \,, \delta g_{0 x} \,,
 1 \, \ma{of} \, \delta g_{0 a} \,, \delta g_{x x} \,,
 1 \, \ma{of} \, \delta g_{x a} \,, 2 \, \ma{of} \, \delta g_{a b} \,, \qquad
 \ma{vector} : 1 \, \ma{of} \, \delta g_{0 a} \,,  
 1 \, \ma{of} \, \delta g_{x a} \,, 1 \, \ma{of} \, \delta g_{a b} \,,
 \end{align}
In total, we have $11 \, (= 4 + 7)$ scalar-type and $4 \, (= 1 + 3)$
 vector-type perturbations.

However, these $15$ degrees of freedom are not independent
because of general covariance and $U(1)$ gauge symmetry. From the
general covariance we have $4 = 3S + 1V$ gauge degrees of freedom and
$4 = 3S + 1V$ constraint equations (Hamiltonian and momentum
constraints). Moreover, the $U (1)$ gauge symmetry removes $1 = 1S$ from
gauge degrees of freedom and $1 = 1S$ from the constraint equation. In
the end, $3 = 11 - 6 - 2$ scalar and $2 = 4 - 2$ vector
perturbations are physical degrees of freedom in this
system.\footnote{In the FLRW case, they correspond to $2$ tensor modes
(gravitational waves), $1$ scalar mode, and $2$ vector modes.}

In linear theory it is useful to perform Fourier decomposition since
each mode evolves independently. In the FLRW case, $3$D rotational
symmetry enables us to take a wave vector to be in the
$x$-direction, ${\bm k} = (k \,, 0 \,, 0)$, for example. On the other
hand, in the current setting, we only have $2$D plane symmetry, which
enables us to take a wave vector to be ${\bm k} = (k_x \,, k_y \,, 0)$,
for example. Hereafter we shall drop the $z$-dependence of perturbations
using this restricted plane symmetry. As a consequence, the scalar
component of a vector $v^a$ is stored in its $y$ component because
$\pa_a v^a = \pa_y v^y$, while the vector component can be read from the
$z$ component of $v^a$.
%%%%%%%%%%%%%%%%%%%%%%%%%%%%%%%%%%%%%%%%%%%%%%%%%%%%%%%%%%%%%%%%%%%%%%%%%%
\subsection{Gauge choice and classification of perturbations}
%%%%%%%%%%%%%%%%%%%%%%%%%%%%%%%%%%%%%%%%%%%%%%%%%%%%%%%%%%%%%%%%%%%%%%%%%%
As discussed in detail in \cite{Watanabe:2010fh}, the most convenient
gauge in studying subhorizon dynamics is the flat gauge, where
information of perturbations is mostly encoded in the matter field
perturbations. Here and hereafter, we shall use the same gauge. In this
gauge, we write the metric perturbations as
\begin{align}
 \delta g_{\mu \nu} =
 \begin{pmatrix}
 - 2 A & e^{2 (\alpha - 2 \beta)} B_x & e^{2 (\alpha + \beta)} B_y & 0 \\
 e^{2 (\alpha - 2 \beta)} B_x & 2 e^{2 (\alpha - 2 \beta)} C & 0 & 0 \\
 e^{2 (\alpha + \beta)} B_y & 0 & 2 e^{2 (\alpha + \beta)} C & 0 \\
 0 & 0 & 0 & - 2 e^{2 (\alpha + \beta)} C
\label{eq:metricpert}
 \end{pmatrix}
 \,,
\end{align}
and the matter perturbations as
\begin{align}
 \delta \phi \,, \qquad
 \delta A_\mu = (\delta A_t \,, 0 \,, \delta A_y \,, 0) \,,
\label{eq:matterpert}
\end{align}
where we have already eliminated three perturbations in spatial
components of the metric and $\delta A_x$ in the matter perturbations
using gauge degrees of freedom.

We insert equations~(\ref{eq:metricpert}) and (\ref{eq:matterpert})
into the action (\ref{action}), and expand it up to the second
order in the perturbations. We give the explicit form in the
appendix. Note that $A$, $B_x$, $B_y$, and $\delta A_t$ are
non-dynamical and can be integrated out because their time derivatives
do not appear in the quadratic action.
%%%%%%%%%%%%%%%%%%%%%%%%%%%%%%%%%%%%%%%%%%%%%%%%%%%%%%%%%%%%%%%%%%%%%%%%%%
\subsection{Slow-roll expansion of the action}
As explained in detail in the appendix, we obtain the action for
canonically normalized perturbation variables after solving the
constraints. We then impose the slow-roll conditions to satisfy current
observational constraints such as a nearly scale-invariant power
spectrum. Here, we simply summarize the action. Our action differs
in detail from that used in the previous work. Bartolo et
al. \cite{Bartolo:2012sd} studied a vector field coupled to inflaton in
the FLRW background, while we study it in the Bianchi type I background.
Watanabe et al. \cite{Watanabe:2010fh} used the Bianchi type I
background, but they used the attractor solution which gives $\dot
I=0$. Our background, which is compatible with the current observational
bound, yields $\dot I/I$ of order the slow-roll parameters (see
appendix~\ref{app:soda}); thus, our action contains terms
proportional to $\dot I$. 

As discussed in the last
subsection, there are three independent dynamical degrees of freedom.
We define canonically normalized variables as
\begin{align}
 \delta \phi ~ \to ~ \vphi \equiv e^{\alpha + \beta} \, \delta \phi \,,
 \qquad C ~ \to ~ \calG \equiv \sqrt{2} e^{\alpha + \beta} \, C \,,
 \qquad \delta A_y ~ \to ~ \calA \equiv f \, \frac{k_x}{\tilde{k}} \,
 \delta A_y
\end{align}
with $\tilde k \equiv \sqrt{k_x^2 + (e^{- 3 \beta} k_y)^2}$. The action of the
scalar perturbations reads
 \begin{align}
 S^\ma{scalar}
 &= \int \dd t \, \frac{\dd^3 k}{(2 \pi)^3} \, e^{\alpha - 2 \beta} \,
 \Bigl( \calL^{\vphi \vphi} + \calL^{\calG \calG} + \calL^{\calA \calA}
 + \calL^{\vphi \calG} + \calL^{\vphi \calA} + \calL^{\calG \calA} \Bigr) \,,
 \end{align}
where the first three and the last three terms correspond to auto- and
cross-correlation terms, respectively. Their explicit forms are given by
\begin{subequations}
\begin{align}  
 \calL^{\vphi \vphi}
 &= \frac{1}{2} |\dot{\vphi}|^2
 - \frac{1}{2} \frac{\tilde{k}^2}{e^{2 (\alpha - 2 \beta)}} |\vphi|^2
 + \frac{1}{2} H^2 \Bigl( 2 + 2 \vep_H + 3 \eta_H
 + \delta m_{\vphi \vphi}^2 \Bigr) |\vphi|^2 \,,
 \label{lag:phiphi} \\
 \calL^{\calG \calG}
 &= \frac{1}{2} |\dot{\calG}|^2
 - \frac{1}{2} \frac{\tilde{k}^2}{e^{2 (\alpha - 2 \beta)}} |\calG|^2
 + \frac{1}{2} H^2 \Bigl[ 2 - \vep_H
 + \calO (\vep_H I)\Bigr] |\calG|^2 \,,
 \label{lag:gg} \\
 \calL^{\calA \calA}
 &= \frac{1}{2} |\dot{\calA}|^2
 - \frac{1}{2} \frac{\tilde{k}^2}{e^{2 (\alpha - 2 \beta)}} |\calA|^2
 + \frac{1}{2} H^2 \left[ 2c \, (2 c - 1)
 + 2c \frac{4 (1 - c) \vep_H + (1 - 4 c) \eta_H}{3}
 + \delta m_{\calA \calA}^2 \right] |\calA|^2 \,,
 \label{lag:AA}   
 \end{align}
 and
 \begin{align}
 \calL^\ma{\vphi \calG}
 &= - 3c H^2 \sqrt{\vep_H} \, I \sin^2 \theta \vphi \, \calG^*
 + (\ma{c.c.}) \,, \\
 \calL^{\vphi \calA}
 &= \sqrt{6} c H \sqrt{I} \sin \theta
 \Bigl( \dot{\vphi} - H \vphi \Bigr) \calA^* + (\ma{c.c.}) \,, \\
 \calL^\ma{\calG \calA}
 &= - \sqrt{\frac{3}{2}} H \sqrt{\vep_H I} \sin \theta
 \Bigl( \dot{\calG} - H \calG \Bigr) \calA^* + (\ma{c.c.})\,,
\end{align}
\end{subequations}
with
\begin{subequations}
\begin{align}
\sin \theta &\equiv \frac{e^{- 3\beta} k_y}{\tilde{k}}\,,\\
 \delta m_{\vphi \vphi}^2 
 &\equiv 2 \Bigl[ 12 c^2 \sin^2 \theta - (4 c^2 + c + 1) \Bigr] I
 + \calO (\vep_H I) \,, \\
 \delta m_{\calA \calA}^2 
 &\equiv - 2c (c + 1) I + \calO (\vep_H I) \,.
\end{align}
\end{subequations}
Here and hereafter, the momentum dependence of the variables is
abbreviated. The slow-roll parameters are defined as
\begin{align}
 \vep_H \equiv - \frac{\dot{H}}{H^2} \,, \qquad
 \eta_H \equiv \frac{\ddot{H}}{2 H \dot{H}} \,.
\end{align}

Equation (\ref{lag:AA}) shows that the effective mass squared of ${\cal
A}$ is $m_\ma{eff}^2 = \frac{1}{2} H^2 \Bigl[ 2c (2c - 1) + \calO (\vep)
\Bigr]$, which is quadratic in $c$. Therefore, two different $c$ can
reproduce the same spectrum of ${\cal A}$. For example, the mass term
vanishes for $c = 0$ because $f (\phi)$ reduces to unity and the action
simply becomes $F_{\mu \nu} F^{\mu \nu}$. The mass term vanishes also
for $c =1/2$ at leading order in the slow-roll expansion. This can be
understood as follows. The action for a homogeneous vector field in the
FLRW universe can be rewritten by introducing a canonical vector field
$\vec{B} = f \vec{A}$ as
\begin{align}
 - \frac{1}{4} \int \dd^4 x \, \sqrt{- g} \, f^2 F_{\mu \nu} F^{\mu \nu}
 & \propto \int \dd^4 x \, e^\alpha f^2 \dot{\vec{A}} \, {}^2 \notag\\
 &= \int \dd^4 x \, e^\alpha \left\{ \dot{\vec{B}} \, {}^2
 + \left[ \frac{1}{f e^\alpha}
 \frac{\dd}{\dd t} (\dot{f} e^\alpha) \right] \vec{B} \, {}^2 \right\} \,.
\end{align}
For $c = 1/2$, $f \propto e^{- 2 \, c \, \alpha} = e^{- \alpha}$; thus,
the second term becomes 
\begin{align}
 \frac{1}{f e^\alpha} \frac{\dd}{\dd t} (\dot{f} e^\alpha) = - \dot{H} =
 \vep_H H^2\,,
\end{align}
i.e., is suppressed by the slow-roll parameter.

%%%%%%%%%%%%%%%%%%%%%%%%%%%%%%%%%%%%%%%%%%%%%%%%%%%%%%%%%%%%%%%%%%%%%%%%%%
%%%%%%%%%%%%%%%%%%%%%%%%%%%%%%%%%%%%%%%%%%%%%%%%%%%%%%%%%%%%%%%%%%%%%%%%%%
\section{Statistical anisotropy}
\label{sec:anisotropy}

\subsection{$g_*$: general potential}
In this section, we shall investigate the statistical properties of
primordial fluctuations generated during anisotropic inflation. To this
end, we use the so-called in-in formalism. In the interaction picture,
the interaction Hamiltonian is given in terms of $\calL^{\vphi \calG}$,
$\calL^{\vphi \calA}$, and $\calL^{\calG \calA}$ as
 \begin{align}
 H^{\vphi \calG}_\ma{int} \equiv \int \dd^3 k \,
 (- e^{\alpha - 2 \beta} \calL^{\vphi \calG}) \,, \quad
 H^{\vphi \calA}_\ma{int} \equiv \int \dd^3 k \,
 (- e^{\alpha - 2 \beta} \calL^{\vphi \calA}) \,, \quad
 H^{\calG \calA}_\ma{int} \equiv \int \dd^3 k \,
 (- e^{\alpha - 2 \beta} \calL^{\calG \calA}) \,.
 \end{align}
The dominant correction to the power spectrum comes from $H^{\vphi
\calA}_\ma{int}$.

We define mode functions and annihilation/creation operators of
perturbations, $\calQ^\lambda$ ($\lambda = \vphi$, $\calG$, $\calA$), as
\begin{align}
 \calQ^\lambda_{\bm k} (\tau)
 &= u^\lambda_{\bm k} (\tau) a^\lambda_{\bm k}
 + u^\lambda_{\bm - k}{}^* (\tau) a^\lambda_{\bm - k}{}^\dagger \,,
\end{align}
where $\tau$ is the conformal time defined as $\dd t = e^{\alpha - 2
\beta} \dd \tau$ and $u^\lambda$ is the solution of the following equation:
\begin{align}
 \pa_\tau^2 u^\lambda_{\bm k}
 + \left( k^2 - \frac{\nu_\lambda^2 - 1/4}{\tau^2} \right) u^\lambda_{\bm k} = 0
 \,,
\end{align}
with $\nu_\lambda$ being
 \begin{subequations}
 \label{indices}
\begin{align}
 \nu_\vphi
 &\equiv \frac{3}{2} + 2 \vep_H + \eta_H + \calO (\vep_H^2 \,, I) \,, \qquad \\
 \nu_\calG
 &\equiv \frac{3}{2} + \vep_H + \calO (\vep_H^2 \,, I) \,, \qquad \\
 \nu_\calA
 &\equiv \frac{3}{2} + \frac{2}{3} \Bigl[ 2 \vep_H - \eta_H + 3 (c-1) \Bigr]
 + \calO (\vep_H^2 \,, I \,, (c-1)^2) \,.
\label{nu:A}
\end{align}
 \end{subequations}
Here and hereafter we neglect the subleading terms of the order of
$\epsilon_H I$ and hence $\tilde{k}$ reduces to $k =
\sqrt{k_x^2+k_y^2}$. We have also used the relation $a = (- k\tau)^{-(1
+ \vep_H)}$, which is valid in the quasi de Sitter approximation; that
is to say, the time dependence of $\vep_H$ is negligible. Using the
Bunch-Davies initial condition, each mode function is given in terms of
the Hankel function, $H_\nu(x)$, as
 \begin{align}
 u^\lambda_{\bm k} = C^\lambda \sqrt{- \tau}
 H_{\nu^\lambda}^{(1)} (- k \tau) \,, \qquad
 C^\lambda = \frac{\sqrt{\pi}}{2} \exp \left[ i \left( \frac{1}{2} \nu^\lambda
 + \frac{1}{4} \right) \pi \right] \,.
\label{eq:mode}
 \end{align}
The power spectrum calculated from the above mode functions must be
consistent with the nearly scale-invariant spectrum, i.e.,
$\nu-3/2={\cal O}(\vep_H,\eta_H)$. This demands $I$ and its
time-variation  be smaller than of order the slow-roll parameters:
\begin{align}
 I \, (t) \leq \calO (\vep_H) \,, \qquad
 \left|\frac{\dot{I}}{H I}\right| \leq \calO
 (\vep_H) ~
 \Longleftrightarrow ~ \left|c - 1\right| \leq \calO (\vep_H) \,.
\end{align}
The first condition is consistent with our starting assumption that $I$ is
small and we can expand the equations in terms of $I$.
The mode function given in equation~(\ref{eq:mode}) is an approximate
solution, which is valid so long as $\nu_\lambda$ is
independent of time. While the variation of the slow-roll parameters is
small, $\nu_\lambda$ may be regarded as constant only during some number of
$e$-folds, $\Delta N$, which is typically of order or less than
$1/\vep_H$. This, however, does not necessarily coincide with the number of
$e$-folds from the horizon exit to the end of inflation. Thus, when we
estimate the evolution of the perturbations on super-horizon scales, the
use of this mode function can be justified only for some number $e$-folds
after the horizon exit. If we were to calculate the full evolution of
the perturbations on super-horizon scales, we would need to perform
numerical calculations without using the Hankel functions as approximate
solutions (which is beyond the scope of this paper). Instead, in the
following analysis, we shall use the Hankel functions as the modes
functions. Thus, our solutions are valid only until some epoch,
$\tau_\vep$, which is of order or less than $1/\vep_H$ $e$-folds after
the horizon exit.

Now we are in the position to calculate $g_*$. We find
\begin{align}
 \frac{\langle \ma{in} | \vphi_{\bm k} \vphi_{\bm p} | \ma{in} \rangle}{\langle
 0 | \vphi_{\bm k} \vphi_{\bm p} | 0 \rangle}  
 &\supset 24 \, \sin^2 \theta \,
 \frac{|C^\vphi|^2 |C^\calA|^2}{(\sqrt{\pi}/2)^4}
 \frac{\Gamma^2 (\nu^\calA)}{\Gamma^2 (\nu^\vphi)} \,
 \left[ \int^{\tau_\vep}_{\tau_l} \dd \tau \,
 a (\tau) H (\tau) \sqrt{I (\tau)} 
 \left( \frac{- k \tau}{2} \right)^{\nu^\vphi - \nu^\calA} \right]^2 \,,
\end{align}
where $\tau_l$ is a certain epoch shortly after the horizon exit,
but well before $\tau_\vep$ at which the Hankel function ceases to
be a good approximation. The contribution from the subhorizon regime,
$\tau<\tau_l$ is negligible because the integrand  oscillates
rapidly and does not contribute to the integral. Using the asymptotic
form of $H_\nu^{(1)} (x)$ in the $x\rightarrow 0$ limit, 
\begin{align}
 H_\nu^{(1)} (x)
 &\to - \frac{i}{\pi} \left( \frac{x}{2} \right)^{- \nu}
 \Bigl[ \Gamma (\nu) + \calO (x^2) \Bigr]
 + \left( \frac{x}{2} \right)^{\nu}
 \left[ \frac{1}{\Gamma (\nu + 1)} + \calO (x^2) \right] \,,
\end{align}
and integrating, we obtain
\begin{align}
 g_*  &\approx - 24 \, I_\he \, 
 \left[ \frac{(- k \tau_\vep)^\delta - (- k \tau_l)^\delta}{
 2^{\nu_\vphi - \nu_\calA} \, \delta} \right]^2
 (-k \tau_\he)^{4(c-1)-2 (\vep_H + \eta_H)}\,,
 \label{result}
\end{align}
with
\begin{align}
\delta \equiv \nu^\vphi - \nu^\calA + \eta_H - 2 (c-1)
 = 2 \left[ \frac{\vep_H + 4 \eta_H}{3} - 2 (c-1) \right] \,.
\end{align}
This expression for $g_*$ is one of the main results of this paper.

If we choose $\tau_\he$ and $\tau_l$ to be at the horizon exit,
$\tau_\ast$, defined by $-k \tau_\ast = 1$ for each mode $\vect k$, the
expression simplifies to
\begin{align}
 g_*  &\approx - 24 \, I_\ast \, 
 \left[ \frac{(- k \tau_\vep)^\delta - 1}
             {2^{\nu_\vphi - \nu_\calA} \, \delta} \right]^2\,.
 \label{result2}
\end{align}
But it should be noted that the final result given above 
 still depends on the choice of $\tau_\vep$. 

Our results on $g_*$ in the new, non-attractor branch of
anisotropic inflation with the leading-order slow-roll terms are
qualitatively different from the previous result given in
equation~(\ref{g:Jiro}), which was based on the attractor branch and
the de Sitter mode function.
Our results explicitly do not depend on 
the number of $e$-folds counted from the end of inflation, $N^2$, but on
$1/\delta^2$. More importantly we have an additional contribution 
 from the numerator in equation~(\ref{result2}), which is
$[(- k \tau_\vep)^\delta - 1]^2$. This main difference
comes from the precise estimation of the integral. By choosing
the time coordinates as $N$ $(\dd N = - H e^\alpha \dd \tau)$ in stead of
$\tau$, we obtain
 \begin{align}
 \int_{\tau_*}^{\tau_\vep} \dd \tau \, (- k \tau)^{- 1 + \delta}  
 \propto \int_{N_\vep}^{N_*} \dd N \, e^{- N \, \delta} \,.
 \end{align}
As $N \approx 60$ and $\delta = \calO (10^{-2})$, the exponent,
$N \delta$, is of order unity, and thus we cannot expand it.
This is the main reason why we have obtained a different formula from
ref.~\cite{Watanabe:2010fh} besides the difference on the background
solution. Of course, if $\delta$ accidentally becomes much smaller than
of order the slow-roll parameter ($\delta \ll 10^{-2}$), we can expand
the exponential factor and the final result may be proportional to
$(\Delta N)^2 = (N_* - N_\vep)^2$. Even in this case, however,
$\Delta N = N_* - N_\vep$ may not be the same as the number of
$e$-folds counted from the end of inflation.
%%%%%%%%%%%%%%%%%%%%%%%%%%%%%%%%%%%%%%%%%%%%%%%%%%%%%%%%%%%%%%%%%%%%%%%%%%
\subsection{$g_*$: $U (\phi) \propto \phi^n$}
\label{subsec:corre}
For slow-roll inflation with a monomial potential, $U (\phi)
\propto\phi^n$, the slow-roll parameters can be expressed in terms of
the number of $e$-folds. Then, it is possible to write $g_*$ in terms of
$N^2_k$, similar to the previous result given in
equation~(\ref{g:Jiro}). However, the resulting $g_*$ is  different
from equation~(\ref{g:Jiro}) quantitatively, as the newly-found factor in
the numerator, $[(- k \tau_\vep)^\delta - 1]^2
 = (1 - e^{- \delta \Delta N})^2$, gives an additional
 correction which is not necessarily of order unity, as shown below.

For $U (\phi)\propto \phi^n$, the slow-roll parameters are given by
 \begin{align}
 \vep_H = \frac{n}{4 N_k} + \calO (\vep_H^2) \,, \qquad
 \eta_H = \frac{2 - n}{4 N_k} + \calO (\vep_H^2) \,.
 \end{align}
Substituting these into the expression of $g_*$, we obtain
\begin{align}
 g_* &= -24 I_* \, N_k^2 \times 2^{- \frac{10 - 3 n}{6 N_k}}
 \left[\frac{1 - e^{- \frac{8 - 3 n}{6} \frac{\Delta N}{N_k}
 + 4 (c-1)\Delta N }}{\frac{8 - 3 n}{6}
 - 4 (c - 1) N_k} \right]^2 \,.
\end{align}
Assuming $\delta \Delta N \ll 1$, this further reduces to
\begin{align}
 g_* & \simeq -24 I_* \, N_k^2 \times
 2^{- \frac{10 - 3 n}{6 N_k}}
 \left( \frac{\Delta N}{N_k} \right)^2 \,.
\end{align}
As is clear from its expression, the dependence on $N_k^2$ can be recovered.
However, at the same time, the additional correction term,
 $2^{- \frac{10 - 3 n}{6 N_k}} (\Delta N/N_k)^2$ appears.

%%%%%%%%%%%%%%%%%%%%%%%%%%%%%%%%%%%%%%%%%%%%%%%%%%%%%%%%%%%%%%%%%%%%%%%%%%
\subsection{Quantum-mechanically generated vector field} 
\label{subsec:QGV}

In this subsection, we will show that anisotropic inflation cannot occur
at high energy scale and then last long enough otherwise too large
statistical anisotropy which contradicts with the current observational
upper bound will be induced due to the quantum-mechanically generated
vector field.

We have adopted perturbative expansion approach in the analysis so far
by splitting the field into that of background and that of fluctuations
around.  Hence there is a definite theoretical limitation for this
approach, that is, if the fluctuations exceeds the background fields,
the analysis is no more trustable.  As we study below, in quasi de
Sitter universe, the expectation value of such quantum-mechanically
generated perturbations at the horizon exit is given by the Hubble
expansion rate at that time. One would expect the expectation
value for the vector field and also its energy density become large if
one consider the mode which exits the horizon earlier because $H$ is a
decreasing function in general.  On the other hand, given the
theoretical prediction for $g_*$ as in equation~(\ref{result2}) and the
observational constraint on it, one can derive the constraint
on the energy density of the background vector field because $I$
can be related with it.  Then using this upper bound on the energy
density of the {\it background} vector field, one can give the
constraint on that of the {\it quantum-mechanically generated} vector
field, which in turn constrains the energy scale during or at the onset
of inflation because the latter should be subdominant comparing to the
former.  In fact combing the results, one finds 
\begin{align}
 \delta \rho^{\vec{A}} \Bigr|_\ma{quantum}
 \leq \, \overline{\rho}^{\vec{A}} \Bigr|_\ma{inf} 
 \simeq 10^{-16} \left( \frac{|g_*|}{10^{- 2}} \right)
 \, \left( \frac{\vep_H H^2}{10^{-12}} \right)_\ma{inf}
 \left[ \frac{(- k \tau_\vep)^\delta - 1}{2^{\nu_\vphi - \nu_\calA} \delta}
 \right]^{-2}.
\label{const:Q}
\end{align}
Strictly speaking, $\epsilon_H$ and $H$ are not constants during
inflation, so their smallest values should be taken for the energy
density of the background vector field.  To this end, let us derive the
explicit form of the energy density of the quantum-mechanically
generated vector field and give the constraint on it.  In passing, it
should be noted that the authors in \cite{Bartolo:2012sd} also
constrained the duration of inflation in this model combining with the
observational upper bound of $g_*$ but based on the previous formula for
$g_*$.

Since the quadratic action for $\calA \simeq f \delta A_y$, which is
related to the fluctuation of the vector field, has already been
obtained in equation~(\ref{lag:AA}), based on the standard quantization
procedure with the choice of the standard Bunch-David vacuum, the
spectrum of the vector field at the horizon exit is found to be
 \begin{align}
 \frac{k^3}{2 \pi^2} \left| \frac{\delta A_y^2}{a^2} \right|
 = \frac{1}{f^2} \left( \frac{H}{2 \pi} \right)^2
 \left( \frac{a H}{k} \right)^{2 \nu_\calA - 3},
 \end{align}
where $a = e^\alpha$ represents a standard scale factor of FLRW universe
and the concrete form of $\nu_\calA$ is presented in
equation~(\ref{nu:A}). Here we keep subleading correction terms,
which avoid the logarithmic divergence in the case of
small-field inflation.  Now the energy density of the vector field,
strictly speaking that of the corresponding electric field will be given
as
\begin{align}
 \delta \rho^{\vec{A}}
 = \left\langle \frac{f^2}{a^2} \delta \dot{A}_y^2 \right\rangle 
 = \int \frac{\dd k}{k} \left( 3 H \times \frac{H}{2 \pi} \right)^2
 \left( \frac{a H}{k} \right)^{2 \nu_\calA - 3},
\end{align}
where $3H$ in the integrand appears due to the time derivative of $\calA$
 and again subdominant contributions are neglected assuming $c \sim 1$.

Unfortunately, it is not an easy task to perform the $k$-integration
 because $H$ and also $\nu_\calA$ in the integrand depend on $k$ and hence one
should take into account those $k$-dependences in the integration.  To
avoid such difficulties, below we shall consider two simple but
interesting cases where one can approximately carry out the integration
with ease.  The first case is a chaotic inflation model where $H$ as
well as $\nu_\calA$ significantly changes.  And the second one is a small
field model where $H$ and $\nu_\calA$ stay almost constant with good accuracy.

 \begin{itemize}
 \item {\it chaotic inflation} \\
In this case, since $H$ significantly changes, a dominant contribution
comes from the time when $H$ takes the largest value, that is, the
onset of inflation, $t_\ma{ini}$.  So a lower-bound is given as
\begin{align}
 \delta \rho^{\vec{A}} (t)
 \geq \left.%  \frac{\Delta k}{k} 
   \left( \frac{3 H^2}{2 \pi} \right)^2
 \left( \frac{a H}{k} \right)^{2 \nu_\calA - 3} \right|_{k = k_\ma{onset}}. 
%  = \left( \frac{3 H^2_\ma{ini}}{2 \pi} \right)^2
%  \left( \frac{a H_\ma{ini}}{k_\ma{ini}} \right)^{2 \nu_\calA - 3} 
\end{align}
Inserting this into equation~(\ref{const:Q}), one gets the constraint
for the Hubble expansion rate at the onset of inflation as
 \begin{align}
 H_\ma{onset}^4
 &\leq 10^{- 16} \, M_\ma{pl}^4
 \left( \frac{|g_*|}{10^{- 2}} \right)
 \left( \frac{\calP_S}{2 \times 10^{- 9}} \right)
 \left( \frac{\vep_H}{10^{-2}} \right)_\ma{CMB}^2
 \left[ \left. \frac{(- k \tau_\vep)^\delta - 1}{2^{\nu_\vphi - \nu_\calA} \delta} 
 \right/ 10 \right]^{- 2},
 \end{align}
where we have introduced the normalization of power spectrum on the CMB scales
 \begin{align}
 \calP_S
 \equiv \left. \frac{1}{2 \vep_H} \left( \frac{H}{2 \pi} \right)^2 
 \right|_\ma{CMB}
 = 2 \times 10^{- 9}, 
 \end{align}
and the subscripts ``onset'' and ``CMB'' represent the onset of
inflation and the CMB scales, respectively. While there is an uncertainty
coming from the last factor in equation~(\ref{const:Q}) as mentioned in
subsection~\ref{subsec:corre}, here we have chosen $10$ as a typical
value.  Now this constraint implies that anisotropic inflation must
start {\it below the GUT scale}.

\item {\it small-field inflation} \\
In this case, the assumption that $H$ and $\nu_\calA$ stay constant gives a
fairly good approximation in reality.  Then one can also explicitly
perform the integration as
\begin{align}
 \delta \rho^{\vec{A}} (t)
 &= \left( \frac{3 H^2}{2 \pi} \right)^2_\ma{inf}
 \int_{k_\ma{ini}}^{aH (t)} \frac{\dd k}{k}
 \left( \frac{a H}{k} \right)^{2 \nu_\calA - 3} 
 = \left( \frac{3 H^2}{2 \pi} \right)^2_\ma{inf}
 \frac{1 - (a H/k_\ma{ini})^{3 - 2 \nu_\calA}}{3 - 2 \nu_\calA}.
\end{align}
Then the constraint of the Hubble expansion rate during inflation is
 obtained as
\begin{align}
 H_\ma{inf}^2
 &\leq 10^{- 8} \, M_\ma{pl}^2 \,
 \left( \frac{|g_*|}{10^{- 2}} \right)
 \left( \frac{\vep_H}{10^{-2}} \right)_\ma{inf}^2
 \left[ \left. \frac{(- k \tau_\vep)^\delta - 1}{2^{\nu_\vphi - \nu_\calA} \delta} 
 \right/ 10 \right]^{- 2}.
\end{align}
This demands that the energy scale of inflation should be
 {\it below the GUT scale}.
\end{itemize}

% Before the end of this section,  we shall note that
%  there might be a possibility some non-perturbative effect
%  might significantly change the above discussion which was
%  based on the perturbative analysis when the quantum-mechanically generated
%  vector field overcomes the background vector field.
% But since it is clearly beyond our scope,
%  we don't consider such possibility here.

%%%%%%%%%%%%%%%%%%%%%%%%%%%%%%%%%%%%%%%%%%%%%%%%%%%%%%%%%%%%%%%%%%%%%%%%%%
%%%%%%%%%%%%%%%%%%%%%%%%%%%%%%%%%%%%%%%%%%%%%%%%%%%%%%%%%%%%%%%%%%%%%%%%%%
\section{Bound on broken rotational invariance during inflation}
\label{sec:rotationalinv}
De Sitter spacetime has ten isometries. Inflation occurs in quasi de
Sitter space, in which time-translation invariance is broken softly by
of order $\vep_H=-\dot H/H^2={\cal O}(10^{-2})$. Broken time-translation
invariance implies broken dilation invariance, which leads to a
departure from the scale-invariant power spectrum of curvature
perturbations of order $\vep_H$. This has been detected convincingly
by WMAP \cite{Hinshaw:2012aka} and Planck \cite{Ade:2013zuv}.

The other nine isometries correspond to three spatial translation and
three rotation, as well as three isometries resulting in the special
conformal transformation in the infinite future
\cite{Creminelli:2012ed}. The latter three isometries are broken during
inflation, while the former six isometries (spatial translation and
rotation) are respected by single-field inflation. Testing these
isometries by the cosmological data thus offers a powerful probe of the
physics of inflation. In this paper, we focus on testing rotational invariance.

In the absence of anisotropic stress, the dimensionless shear due to anisotropic
expansion, $\dot\beta/H$, decays as $a^{-3}$ because the Einstein
equation gives $\ddot{\beta} + 3 H \dot{\beta}=0$ (see
equation~(\ref{eq:beta})). If the universe had the shear of order unity
before inflation, and the total number of $e$-folds of inflation was
just enough to solve the flatness problem, the natural value for the
residual shear is $|\dot\beta|/H=e^{-3N}\approx e^{-180}$. 

On the other
hand, Maleknejad and Sheikh-Jabbari \cite{Maleknejad:2012as} showed
that it is possible to generate $\dot\beta/H$ of order $\vep_H$ or
smaller while not spoiling slow-roll inflation, if there is a persisting
anisotropic stress during inflation. This is the scenario that we studied
in this paper. Therefore, without fine-tuning, we would expect
$\dot\beta/H$ to be either $e^{-180}$ or $10^{-2}$. Using $I\approx{\cal
V}/(\vep_HU)$, $|\dot\beta|/H\approx {\cal V}/(3U)$, and
 $g_*\approx -24I\vep_H^{- 2}$ (up to corrections of order unity),
 we find $g_*\approx -72 \, \vep_H^{- 3} \, (|\dot\beta|/H)$. 
Therefore, we would expect $|g_*|$ to
be either $10^{-70}$ or $10^6$ for $\vep_H \approx 10^{-2}$. The latter is
excluded by the Planck data, which gives the $95\%$ bound of
$|g_*|<0.03$. Reversing the argument, we find a bound on the shear
during inflation of order
\begin{equation}
 \frac{|\dot\beta|}{H}<10^{-9}.
\end{equation}
This bound is another main result of this paper.
This result relies on the specific model of anisotropic inflation
connecting $g_*$ to the background spacetime. Generality of this bound
is not known at the moment, and it merits further study, perhaps in the
language of Effective Field Theory (EFT) of inflation
\cite{Cheung:2007st,Cannone:2014uqa}. In the nominal case of
single-field inflation, a Nambu-Goldstone boson associated with breaking
of time-translation invariance gives the curvature perturbation. In the
scenario we studied in this paper, spatial-rotation invariance is broken
such that $O(3)\to O(2)$. This would generate two additional
Nambu-Goldstone bosons, disguising themselves as, e.g., anisotropic stresses,
$\bigl( \pi^{\vec{A}} \bigr)^x{}_x$ and $\bigl( \pi^{\vec{A}}
\bigr)^y{}_y$. Developing EFT for this case would shed light on
generality of the connection between $g_*$ and rotational symmetry of
the background spacetime.

%%%%%%%%%%%%%%%%%%%%%%%%%%%%%%%%%%%%%%%%%%%%%%%%%%%%%%%%%%%%%%%%%%%%%%%%%%
%%%%%%%%%%%%%%%%%%%%%%%%%%%%%%%%%%%%%%%%%%%%%%%%%%%%%%%%%%%%%%%%%%%%%%%%%%
\section{Conclusion}
\label{sec:conclusion}

In this paper, we have used the observational bound on statistical
anisotropy in the power spectrum parametrized by $g_*$
\cite{Kim:2013gka} to infer the bound on broken rotational invariance in
the background spacetime during inflation. We found a rather stringent
(albeit model-dependent) bound, $|\dot\beta|/H<10^{-9}$, whereas the
natural expectation would be either $e^{-3N}\approx e^{-180}$ or
$\vep_H\approx 10^{-2}$.

To derive this bound, we uncovered two subtleties missing in the
previous calculations of $g_*$ from anisotropic inflation with
$f^2(\phi)F^2$ in the action \cite{Watanabe:2009ct}. One is that the
attractor solution giving $I=(c-1)/c^2$ is not compatible with the
observational bound on $g_*$ except for the very special case with
fine-tuning, and thus another branch is discussed in this paper. The
other is that the leading slow-roll corrections to the de Sitter mode
function changes the prediction for $g_*$ qualitatively, by replacing
$g_*\propto N_k^2$ with $[2(\vep_H+4\eta_H)/3-4(c-1)]^{-2}$. For
large-field inflation models with a monomial potential, our result gives
an order-unity correction to the previous formula for $g_*$, i.e.,
$g_*=-{\cal O}(1)\times 24IN_k^2$, because $N_k={\cal O}(1)/\vep_H$.
For small-field models, the slow-roll parameters are not related
with $N_k$.  It should be noted that in this paper we have derived
the theoretical prediction for $g_*$ after some number $e$-folds after
the horizon exit, which is in general not the end of inflation.  And
hence to predict its final value of $g_*$ at the end of inflation, the
superhorizon evolution of it must be taken into account which is beyond
our current scope.  We hope this issue will also be reexamined in near
future.

The connection between $g_*$  and the background spacetime depends on
models, and generality of our result is still unclear. To address this,
EFT of inflation breaking spatial rotation symmetry seems a promising avenue.
%%%%%%%%%%%%%%%%%%%%%%%%%%%%%%%%%%%%%%%%%%%%%%%%%%%%%%%%%%%%%%%%%%%%%%%%%%
%%%%%%%%%%%%%%%%%%%%%%%%%%%%%%%%%%%%%%%%%%%%%%%%%%%%%%%%%%%%%%%%%%%%%%%%%%
\begin{acknowledgments}
A.N. would like to thank Nicola Bartolo, Ryo Namba, Misao Sasaki and Jiro Soda
 for fruitful discussions. M.Y. would also like to thank Jiro Soda for useful
 discussion. A.N. would like to thank Max-Planck-Institut f\"ur
 Astrophysik (MPA), APC (CNRS-Universit\`e Paris 7), INFN, Sezione di Padova,
 where this work was advanced, for warm hospitality. 
 A.N. would also like to thank the Yukawa Institute for Theoretical Physics
 at Kyoto University for hospitality as well as the YITP Workshop YITP-X-13-03
 for useful discussions during the workshop.
 A.N. was supported in part by JSPS Postdoctoral Fellowships
 for Research Abroad and Grant-in-Aid for JSPS Fellows No. 26-3409.  
 M.Y. was in part supported by the JSPS
 Grant-in-Aid for Scientific Research Nos. 25287054 and 26610062.
\end{acknowledgments}
%%%%%%%%%%%%%%%%%%%%%%%%%%%%%%%%%%%%%%%%%%%%%%%%%%%%%%%%%%%%%%%%%%%%%%%%%%
%%%%%%%%%%%%%%%%%%%%%%%%%%%%%%%%%%%%%%%%%%%%%%%%%%%%%%%%%%%%%%%%%%%%%%%%%%
%%%%%%%%%%%%%%%%%%%%%%%%%%%%%%%%%%%%%%%%%%%%%%%%%%%%%%%%%%%%%%%%%%%%%%%%%%
\appendix
\section{Detailed derivation of the reduced action}
In this appendix, we explain in detail how to obtain the reduced
action for the canonical variables under the slow-roll approximation.
Essential steps are :
\begin{itemize}
 \item integrate out non-dynamical variables, such as 
 the lapse function and shift vector.
 \item introduce canonical variables so as to normalize the kinetic term
 \item perform the slow-roll expansion
\end{itemize}
%%%%%%%%%%%%%%%%%%%%%%%%%%%%%%%%%%%%%%%%%%%%%%%%%%%%%%%%%%%%%%%%%%%%%%%%%%
\subsection{Definitions and expansion of the action}
First we summarize the definition of perturbations. The metric
perturbations are given by
\begin{align}
 \delta g_{\mu \nu} =
 \begin{pmatrix}
 - 2 A & e^{2 (\alpha - 2 \beta)} B_x & e^{2 (\alpha + \beta)} B_y & 0 \\
 e^{2 (\alpha - 2 \beta)} B_x & 2 e^{2 (\alpha - 2 \beta)} C & 0 & 0 \\
 e^{2 (\alpha + \beta)} B_y & 0 & 2 e^{2 (\alpha + \beta)} C & 0 \\
 0 & 0 & 0 & - 2 e^{2 (\alpha + \beta)} C
 \end{pmatrix}
 \,,
\end{align}
and matter perturbations are
\begin{align}
 \delta \phi \,, \qquad
 \delta A_\mu = (\delta A_t \,, 0 \,, \delta A_y \,, 0) \,.
\end{align}
After substituting the above perturbations into the action and expanding
it up to second order in perturbations, one obtains
\begin{align}
 S^\ma{scalar}
 &= \int \dd x^4 \sqrt{- g} \left[ \frac{1}{2} R
 - \frac{1}{2} g^{\mu \nu} \pa_\mu \phi \pa_\nu \phi - U (\phi)
 - \frac{1}{4} f^2 (\phi) F_{\mu \nu} F^{\mu \nu} \right]^\ma{scalar} \notag\\ 
 &= \int \dd x^4 \, e^{3 \alpha} 
 \left\{ - 3 \left( A^2 - A C + \frac{1}{2} C^2 \right) (H^2 - \dot{\beta}^2)
 \right. \notag\\
 & \qquad 
 - (A - C) \Bigl[ (2 H - \dot{\beta}) B_{y, y}
 - 2 (H + \dot{\beta}) (\dot{C} - B_{x, x}) \Bigr]
 + 2 (H + \dot{\beta}) (B_x C_{, x} + B_y C_{, y}) \notag\\
 & \qquad 
 + \frac{1}{4} \Bigl( e^{- 6 \beta} B_{x, y}^2 + 2 B_{x, y} B_{y, x} 
 - 4 B_{x, x} B_{y, y} + e^{6 \beta} B_{y, x}^2 \Bigr)
 - e^{- 2 (\alpha - 2 \beta)} C_{, x}^2
 - e^{- 2 (\alpha + \beta)} C_{, y}^2 + \dot{C}^2 \notag\\
 & \qquad  + \frac{1}{2} \frac{f^2 \dot{u}^2}{e^{2 (\alpha - 2 \beta)}} \left[
 \left( \frac{f_{\phi \phi}}{f} + \frac{f_\phi^2}{f^2} \right) \delta \phi^2
 - 2 \frac{f_\phi}{f} \delta \phi (A + C)
 + A^2 + AC + \frac{1}{2} C^2 \right] \notag\\
 & \qquad 
 + \frac{f^2 \dot{u}}{e^{2 (\alpha - 2 \beta)}} \left[ 
 - \left( 2 \frac{f_\phi}{f} \delta \phi - A - C \right) \delta A_{t, x} 
 + \delta A_{y, x} B_y \right] \notag\\
 & \qquad 
 + \frac{1}{2} \frac{f^2}{e^{2 (\alpha - 2 \beta)}} \Bigl[ \delta A_{t, x}^2
 + e^{- 6 \beta} (\dot{\delta A}_y - \delta A_{t, y})^2
 - e^{- 2 (\alpha + \beta)} \delta A_{y, x}^2 \Bigr] \notag\\
 & \qquad 
 + \frac{1}{2} \left[ \dot{\delta \phi}^2
 - 2 \dot{\phi} (A - C) \dot{\delta \phi}
 + \dot{\phi}^2 \left( A^2 - A C + \frac{C^2}{2} \right) \right]
 - C \left( A + \frac{C}{2} \right) U
 - (A + C) U_\phi \delta \phi \notag\\
 & \qquad \left.
 - \dot{\phi} (B_x \delta \phi_{, x} + B_y \delta \phi_{, y})
 - \frac{1}{2} \Bigl( e^{- 2 (\alpha - 2 \beta)} \delta \phi_{, x}^2
 + e^{- 2 (\alpha + \beta)} \delta \phi_{, y}^2 \Bigr)
 - \frac{1}{2} U_{\phi \phi} \delta \phi^2 \right\}.
 \label{expansion:action}
\end{align}
%%%%%%%%%%%%%%%%%%%%%%%%%%%%%%%%%%%%%%%%%%%%%%%%%%%%%%%%%%%%%%%%%%%%%%%%%%
\subsection{Solving constraints}
Since $A$, $B_x$, $B_y$, and $\delta A_t$ are non-dynamical (their time
derivatives do not appear in the action), they can be integrated out.
The terms related to $\delta A_t$ are given by
\begin{align}
 S^\ma{scalar}
 &\supset \int \dd x^4 \, e^{\alpha} f^2 \left[ - e^{4 \beta}
 \left( 2 \frac{f_\phi}{f} \delta \phi - A - C \right) \dot{u} \delta A_{t, x}
 + \frac{1}{2} e^{4 \beta} \delta A_{t, x}^2
 + \frac{1}{2} e^{- 2 \beta} (\delta A_{t, y}^2
 - 2 \dot{\delta A_y} \delta A_{t, y}) \right] \,.
 \end{align}
Completing the square yields
 \begin{align}
 & \int \dd t \, \frac{\dd^3 k}{(2 \pi)^3} \, e^{3 \alpha}
 \left[ \frac{1}{2} \left| \delta A_t + i \frac{k_x}{\tilde{k}^2}
 \left( 2 \frac{f_\phi}{f} \delta \phi - A - C \right) \dot{u}
 + i \frac{k_y}{\tilde{k}^2} e^{- 6 \beta} \dot{\delta A_y}
 \right|^2 \right. \notag\\
 & \qquad \qquad \qquad \qquad \qquad \left.
 - \frac{1}{2} \left| \frac{k_x}{\tilde{k}^2}
 \left( 2 \frac{f_\phi}{f} \delta \phi - A - C \right) \dot{u}
 + \frac{k_y}{\tilde{k}^2} e^{- 6 \beta} \dot{\delta A_y} \right|^2 \right] \,,
 \end{align}
where we have moved to Fourier space and defined $\tilde{k}$ as
 \begin{align}
 \tilde{k} \equiv \sqrt{k_x^2 + (e^{- 3 \beta} k_y)^2} \,.  
 \end{align}
Note that each perturbation variable depends on momenta, although we do
not write it explicitly. 

In a similar way, the terms containing $B_x$ are given by
\begin{align}
 & \int \dd^4 x \, e^{3 \alpha}
 \biggl[ 2 (H + \dot{\beta}) \Bigl( B_x C_{, x} - (A - C) B_{x, x} \Bigr)
 \notag\\
 & \qquad \qquad \qquad \left. 
 + \frac{1}{4} \Bigl( e^{- 6 \beta} B_{x, y}^2 + 2 B_{x, y} B_{y, x} 
 - 4 B_{x, x} B_{y, y} \Bigr) - \dot{\phi} B_x \delta \phi_x \right] \,,
 \end{align}
which can be rewritten as
 \begin{align}
 & \int \dd t \, \frac{\dd^3 k}{(2 \pi)^3} \, e^{3 \alpha} \left[
 e^{- 6 \beta} k_y^2 \left| \frac{1}{2} B_x
 + \frac{k_x}{k_y^2} e^{6 \beta} \left( i \, 2 (H + \dot{\beta}) A
 - \frac{1}{2} k_y B_y - i \dot{\phi} \delta \phi \right) \right|^2
 \right. \notag\\
 & \qquad \qquad \qquad \qquad \qquad \left.
 - \frac{1}{4} k_x^2 e^{6 \beta} \left| B_y
 - i \frac{2}{k_y} \Bigl( 2 (H + \dot{\beta}) A
 - \dot{\phi} \delta \phi \Bigr) \right|^2 \right] \,.
 \end{align}
As for $B_y$, we find that $B_y$ appears only linearly in the action:
\begin{align}
 & \int \dd t \, \frac{\dd^3 k}{(2 \pi)^3} \, e^{3 \alpha} \,
 i k_y B_y^* \left[ \left( 2 H - \dot{\beta} 
 + 2 (H + \dot{\beta}) \frac{k_x^2}{k_y^2} e^{6 \beta} \right) A
 + 3 \dot{\beta} C \right. \notag\\
 & \qquad \qquad \qquad \qquad \qquad \qquad \left.
 + \frac{f^2}{e^{2 (\alpha - 2 \beta)}} \frac{k_x}{k_y} \dot{u} \delta A_y
 - \left( 1 + \frac{k_x^2}{k_y^2} e^{6 \beta} \right)
 \dot{\phi} \delta \phi \right]\,,
 \end{align}
and thus $B_y$ behaves like a Lagrange multiplier. Variation with
respect to $B_y$ allows us to express $A$ in terms of the other dynamical
variables as
\begin{align}
 A
 = \frac{1}{\lambda} \left( - 3 \dot{\beta} C
 - \frac{f^2}{e^{2 (\alpha - 2 \beta)}} \frac{k_x}{k_y} \dot{u} \delta A_y
 + \frac{\tilde{k}^2}{k_y^2} e^{6 \beta} \dot{\phi} \delta \phi \right)
\end{align}
with
\begin{align}
 \lambda \equiv
 H - 2 \dot{\beta} + (H + \dot{\beta}) \left( 1
 + 2 \frac{k_x^2}{k_y^2} e^{6 \beta} \right) \,.
\end{align}

It is now manifest that the action can be written only in terms of the
dynamical variables $C$, $\delta \phi$, and $\delta A_y$. The kinetic
terms in the action take the following form,
 \begin{align}
 S^\ma{scalar}
 &\supset \int \dd t \, \frac{\dd k^3}{(2 \pi)^3} \, e^{\alpha - 2 \beta}
 \left( e^{2 (\alpha + \beta)} |\dot{C}|^2
 + \frac{1}{2} e^{2 (\alpha + \beta)} |\dot{\delta \phi}|^2
 + \frac{1}{2} f^2 \frac{k_x^2}{\tilde{k}^2} |\dot{\delta A_y}|^2 \right) \,,
 \end{align}
which enables us to define canonically normalized variables as
\begin{align}
 C \to \calG \equiv \sqrt{2} e^{\alpha + \beta} C \,, \qquad
 \phi \to \vphi \equiv e^{\alpha + \beta} \delta \phi \,, \qquad
 \delta A_y \to \calA \equiv f \frac{k_x}{\tilde{k}} \delta A_y \,.
\end{align}
%%%%%%%%%%%%%%%%%%%%%%%%%%%%%%%%%%%%%%%%%%%%%%%%%%%%%%%%%%%%%%%%%%%%%%%%%%
\subsection{Action written with canonical variables}
We shall write down the reduced action in terms of the canonical
variables $\calG$, $\vphi$, and $\calA$. 
 \begin{align}   
 S^\ma{scalar}
 &= \int \dd t \, \frac{\dd k^3}{(2 \pi)^3} \, e^{\alpha - 2 \beta}
 \Bigl( \calL^{\calG \calG} + \calL^{\vphi \vphi} + \calL^{\calA \calA}
 + \calL^{\calG \vphi} + \calL^{\vphi \calA} + \calL^{\calA \calG} \Bigr)\,,
 \end{align}
where
\begin{align}
 \calL^{\calG \calG}
 &= \frac{1}{2} |\dot{\calG}|^2
 - \frac{1}{2} \frac{\tilde{k}^2}{e^{2 (\alpha - 2 \beta)}} |\calG|^2
 + \frac{1}{2} \left\{ \frac{1}{e^{\alpha - 2 \beta}} \frac{\dd}{\dd t} \left[
 e^{\alpha - 2 \beta} (H + \dot{\beta}) \left( 2
 + 3 \frac{\dot{\beta}}{\lambda} \right) \right]
 + 3 (H + \dot{\beta})^2 \left( 1 + 2 \frac{\dot{\beta}}{\lambda} \right)
 \right\} |\calG|^2 \notag\\
 & \qquad 
 - \frac{1}{2} \left\{ \left( 1 + 9 \frac{\dot{\beta}^2}{\lambda^2} \right) U
 + 36 e^{6 \beta} \frac{k_x^2}{k_y^2} (H + \dot{\beta})^2
 \frac{\dot{\beta}^2}{\lambda^2}
 + \frac{f^2 \dot{u}^2}{2 e^{2 (\alpha - 2 \beta)}}
 \left[ \frac{k_x^2}{\tilde{k}^2}
 \left( 1 - 3 \frac{\dot{\beta}}{\lambda} \right)^2
 + 6 \frac{\dot{\beta}}{\lambda} \right] \right\} |\calG|^2 \,, 
\end{align}
\begin{align}
 \calL^{\vphi \vphi}
 &= \frac{1}{2} |\dot{\vphi}|^2
 - \frac{1}{2} \frac{\tilde{k}^2}{e^{2 (\alpha - 2 \beta)}} |\vphi|^2
 + \frac{1}{2} \frac{f^2 \dot{u}^2}{e^{2 (\alpha - 2 \beta)}}
 \left[ \frac{f_{\phi \phi}}{f} + \frac{f_\phi^2}{f^2}
 - 2 \frac{f_\phi}{f} e^{6 \beta} \frac{\tilde{k}^2}{k_y^2}
 \frac{\dot{\phi}}{\lambda}
 - \frac{k_x^2}{\tilde{k}^2} \left( 2 \frac{f_\phi}{f}
 - e^{6 \beta} \frac{\tilde{k}^2}{k_y^2} \frac{\dot{\phi}}{\lambda} \right)^2
 \right] |\vphi|^2 \notag\\
 & \qquad 
 + \frac{1}{2} \left[ (2 H - \dot{\beta}) (H + \dot{\beta})
 + 3 H^2 e^{6 \beta} \frac{\tilde{k}^2}{k_y^2} \frac{\dot{\phi}^2}{\lambda}
 + \dot{H} + \ddot{\beta}
 + e^{6 \beta} \frac{\tilde{k}^2}{k_y^2} \frac{\dot{\phi} \ddot{\phi}}{\lambda} 
 + \dot{\phi} \frac{\dd}{\dd t} \left( e^{6 \beta} \frac{\tilde{k}^2}{k_y^2}
 \frac{\dot{\phi}}{\lambda} \right) \right] |\vphi|^2 \notag\\
 & \qquad 
 - \left[ \left( e^{6 \beta} \frac{\tilde{k}^2}{k_y^2}
 \frac{\dot{\phi}}{\lambda} \right)^2 U
 + e^{6 \beta} \frac{\tilde{k}^2}{k_y^2} \frac{\dot{\phi}}{\lambda} U_\phi
 + \frac{1}{2} U_{\phi \phi} + e^{6 \beta} \frac{k_x^2}{k_y^2} \dot{\phi}^2
 \left( 1 - 2 e^{6 \beta} \frac{\tilde{k}^2}{k_y^2}
 \frac{H + \dot{\beta}}{\lambda} \right)^2 \right] |\vphi|^2 \,, 
\end{align}
\begin{align}
 \calL^{\calA \calA}
 &= \frac{1}{2} |\dot{\calA}|^2
 - \frac{1}{2} \frac{\tilde{k}^2}{e^{2 (\alpha - 2 \beta)}} |\calA|^2
 + \frac{1}{2} \left\{ \frac{1}{e^{\alpha - 2 \beta}} \frac{d}{d t} \left[
 e^{\alpha - 2 \beta} \left( \frac{\dot{f}}{f}
 - \frac{\dot{\tilde{k}}}{\tilde{k}}
 + \frac{f^2 \dot{u}^2}{e^{2 (\alpha - 2 \beta)} \lambda} \right) \right]
 + \left( \frac{\dot{f}}{f} - \frac{\dot{\tilde{k}}}{\tilde{k}} \right)^2
 \right\} |\calA|^2 \notag\\
 & \qquad
 + \frac{f^2 \dot{u}^2}{e^{2 (\alpha - 2 \beta)} \lambda} \left(
 \frac{\dot{f}}{f} - \frac{\dot{\tilde{k}}}{\tilde{k}}
 - 4 e^{12 \beta} \frac{k_x^2 \tilde{k}^2}{k_y^4}
 \frac{(H + \dot{\beta})^2}{\lambda}
 - e^{6 \beta} \frac{\tilde{k}^2}{k_y^2} \frac{U}{\lambda}
 - \frac{1}{2} \frac{f^2 \dot{u}^2}{e^{2 (\alpha - 2 \beta)} \lambda}
 e^{6 \beta} \frac{k_x^2}{k_y^2} \right) |\calA|^2 \,,
 \end{align}
 and
 \begin{align}
 2 \calL^{\calG \vphi}
 &= \sqrt{2} (H + \dot{\beta}) e^{6 \beta} \frac{\tilde{k}^2}{k_y^2}
 \frac{\dot{\phi}}{\lambda} \, \vphi \, \dot{\calG}^* 
 + \frac{1}{\sqrt{2}} \left( 1 + 3 \frac{\dot{\beta}}{\lambda} \right)
 \dot{\phi} \, \dot{\vphi} \, \calG^* \notag\\
 & \qquad
 - \frac{1}{\sqrt{2}} \frac{f^2 \dot{u}^2}{e^{2 (\alpha - 2 \beta)}}
\left( 1 - 3 \frac{\dot{\beta}}{\lambda} \right)
 \left[ \frac{f_\phi}{f} - \frac{k_x^2}{\tilde{k}^2}
 \left( 2 \frac{f_\phi}{f} - e^{6 \beta} \frac{\tilde{k}^2}{k_y^2}
 \frac{\dot{\phi}}{\lambda} \right) \right] \vphi \, \calG^* \notag\\
 & \qquad 
 - \frac{1}{\sqrt{2}} \left[ \left( 1 + 3 \frac{\dot{\beta}}{\lambda}
 \right) \dot{\phi} (H + \dot{\beta})
 + \left( 2 (H + \dot{\beta})^2 - 6 \frac{\dot{\beta}}{\lambda} U
 - \frac{f^2 \dot{u}^2}{e^{2 (\alpha - 2 \beta)}} \right)
 e^{6 \beta} \frac{\tilde{k}^2}{k_y^2} \frac{\dot{\phi}}{\lambda}
 \right] \vphi \, \calG^* \notag\\
 & \qquad
 - \frac{1}{\sqrt{2}} \left[
 \left( 1 - 3 \frac{\dot{\beta}}{\lambda} \right) U_\phi
 + 12 e^{6 \beta} \frac{k_x^2}{k_y^2}
 (H + \dot{\beta}) \frac{\dot{\beta}}{\lambda}
 \dot{\phi} \left( 1 - 2 e^{6 \beta} \frac{\tilde{k}^2}{k_y^2}
 \frac{H + \dot{\beta}}{\lambda} \right) \right] \vphi \, \calG^*
 + (c.c.) \,, 
\end{align}
\begin{align}
 2 \calL^{\vphi \calA}
 &= \frac{f \dot{u}}{e^{\alpha - 2 \beta} \lambda} e^{3 \beta}
 \frac{\tilde{k}}{k_y} \left[ \dot{\phi} \, \dot{\vphi} \, \calA^*
 - e^{- 6 \beta} \frac{k_y^2}{\tilde{k}^2}
 \left( 2 \lambda \frac{f_\phi}{f} - e^{6 \beta} \frac{\tilde{k}^2}{k_y^2}
 \dot{\phi} \right) \vphi \, \dot{\calA}^* 
 + \left( 2 e^{6 \beta} \frac{\tilde{k}^2}{k_y^2}
 \frac{\dot{\phi}}{\lambda} U - \dot{\phi} (H + \dot{\beta}) + U_\phi \right)
 \, \vphi \, \calA^* \right] \notag\\
 & \qquad 
 + \frac{f \dot{u}}{e^{\alpha - 2 \beta}} e^{3 \beta} \frac{\tilde{k}}{k_y}
 \left[ - 4 e^{6 \beta} \frac{k_x^2}{k_y^2}
 \frac{(H + \dot{\beta}) \dot{\phi}}{\lambda}
 \left( 1 - 2 e^{6 \beta} \frac{\tilde{k}^2}{k_y^2}
 \frac{H + \dot{\beta}}{\lambda} \right) 
 + \frac{f^2 \dot{u}^2}{e^{2 (\alpha - 2 \beta)} \lambda} \frac{f_\phi}{f} \right]
 \vphi \, \calA^* \notag\\
 & \qquad 
 + \frac{f \dot{u}}{e^{\alpha - 2 \beta}} e^{3 \beta} \frac{\tilde{k}}{k_y}
 \left[ e^{- 6 \beta} \frac{k_y^2}{\tilde{k}^2}
 \left( \frac{\dot{f}}{f} - \frac{\dot{\tilde{k}}}{\tilde{k}} \right) 
 - \frac{f^2 \dot{u}^2}{e^{2 (\alpha - 2 \beta)} \lambda}
 \frac{k_x^2}{\tilde{k}^2} \right]
 \left( 2 \frac{f_\phi}{f} - e^{6 \beta} \frac{\tilde{k}^2}{k_y^2}
 \frac{\dot{\phi}}{\lambda} \right) \vphi \, \calA^*
 + (c.c.) \,, 
\end{align}
\begin{align}
 2 \calL^{\calA \calG}
 &= \frac{f \dot{u}}{e^{\alpha - 2 \beta}} e^{3 \beta} \frac{\tilde{k}}{k_y}
 \left[ \frac{1}{\sqrt{2}} e^{- 6 \beta} \frac{k_y^2}{\tilde{k}^2}
 \left( 1 - 3 \frac{\dot{\beta}}{\lambda} \right) \calG \, \dot{\calA}^*
 - \sqrt{2} \frac{H + \dot{\beta}}{\lambda} \, \dot{\calG} \, \calA^* \right]
 \notag\\
 & \qquad
 + \sqrt{2} \frac{f \dot{u}}{e^{\alpha - 2 \beta} \lambda}
 e^{3 \beta} \frac{\tilde{k}}{k_y}
 \left[ (H + \dot{\beta})^2 \left( 1 - 12 e^{6 \beta} \frac{k_x^2}{k_y^2} 
 \frac{\dot{\beta}}{\lambda} \right) - 3 \frac{\dot{\beta}}{\lambda} U
 - \frac{1}{2} \frac{f^2 \dot{u}^2}{e^{2 (\alpha - 2 \beta)}} \right]
 \calG \, \calA^* \notag\\
 & \qquad 
 - \frac{1}{\sqrt{2}} \frac{f \dot{u}}{e^{\alpha - 2 \beta}}
 e^{- 3 \beta} \frac{k_y}{\tilde{k}}
 \left( 1 - 3 \frac{\dot{\beta}}{\lambda} \right)
 \left( \frac{\dot{f}}{f} - \frac{\dot{\tilde{k}}}{\tilde{k}}
 - e^{6 \beta} \frac{k_x^2}{k_y^2}
 \frac{f^2 \dot{u}^2}{e^{2 (\alpha - 2 \beta)} \lambda} \right) \calG \, \calA^*
 + (c.c.) \,.
 \end{align}
%%%%%%%%%%%%%%%%%%%%%%%%%%%%%%%%%%%%%%%%%%%%%%%%%%%%%%%%%%%%%%%%%%%%%%%%%%
%%%%%%%%%%%%%%%%%%%%%%%%%%%%%%%%%%%%%%%%%%%%%%%%%%%%%%%%%%%%%%%%%%%%%%%%%%
\section{Slow-roll expansion}
%%%%%%%%%%%%%%%%%%%%%%%%%%%%%%%%%%%%%%%%%%%%%%%%%%%%%%%%%%%%%%%%%%%%%%%%%%
\subsection{Relation among small quantities in the slow-roll approximation}
Since we are interested in the leading-order terms in the slow-roll
expansion, we try to express small quantities in terms of the slow-roll
parameters, $\vep_H$ and $\eta_H$, defined by
 \begin{align}
 \vep_H \equiv - \frac{\dot{H}}{H^2} \,, \qquad
 \eta_H \equiv \frac{\ddot{H}}{2 H \dot{H}} \,,
 \end{align}
 and the additional anisotropy parameter
 \begin{align}
 I \equiv 2 \frac{U}{U_\phi^2} \calV \,, \qquad
 \calV \equiv \frac{f^2 \dot{u}^2}{e^{2 (\alpha - 2 \beta)}} 
 = \frac{C_A^2}{f^2 e^{4 (\alpha + \beta)}} \,.
\label{def:I}
 \end{align}

The equation for $\beta$ reads
\begin{align}
 \ddot{\beta} + 3 H \dot{\beta} = \frac{1}{3} \calV \quad \to \quad 
 \dot{\beta}
 \approx \frac{1}{3} e^{- 3 \alpha} \int \dd t \, e^{3 \alpha}
 \left( \frac{1}{2} \frac{U_\phi^2}{U^2} \right) U I.
\end{align}
The second Friedmann equation gives
\begin{align}
 \dot{H} + 3 H^2 = U + \frac{1}{6} \calV \quad \to \quad
 U = \frac{3 - \vep_H}{1 + \frac{1}{12} \frac{U_\phi^2}{U^2} I} H^2.
\end{align}
Combining this equation with the first Friedmann equation yields
\begin{align}
 \frac{1}{2} \frac{\dot{\phi}^2}{H^2}
 = \vep_H - \left( \frac{1}{2} \frac{U_\phi^2}{U^2} \right)
 \left( 1 - \frac{1}{3} \vep_H \right) I + \calO (I^2).
\end{align}
The scalar field equation gives the following relation,
\begin{align}
 \ddot{\phi} + 3 H \dot{\phi} + U_\phi (1 - c I) = 0 \quad \to \quad
 \sqrt{\frac{\frac{1}{2} \frac{\dot{\phi}^2}{H^2}}{\frac{1}{2}
 \frac{U_\phi^2}{U^2}}} 
 = \frac{3 - \vep_H}{3 + \frac{\ddot{\phi}}{H \dot{\phi}}}
 \left[ 1 - \left( c + \frac{1}{12} \frac{U_\phi^2}{U^2} \right) I \right]
 + \calO (I^2)\,.
\end{align}
Taking a time derivative of the second Friedmann equation gives
\begin{align}
 \eta_H 
 = \frac{\frac{1}{2} \frac{\dot{\phi}^2}{H^2}}{\vep_H}
 \frac{\ddot{\phi}}{H \dot{\phi}}
 + 2 \frac{\frac{1}{2} \frac{U_\phi^2}{U^2}}{\vep_H}
 \left( 1 - \frac{1}{3} \vep_H \right)  
 \left( c \sqrt{\frac{\frac{1}{2} \frac{\dot{\phi}^2}{H^2}}{\frac{1}{2}
 \frac{U_\phi^2}{U^2}}} - 1 \right) I + \calO (I^2)\,.
\end{align}

Finally, by rewriting $\dot{f}/f$ as
\begin{align}
 \frac{\dot{f}}{f} = \frac{f_\phi}{f} \dot{\phi}
 = 2 c \frac{U}{U_\phi} \dot{\phi} \,,
\end{align}
 one finds the following expression,
 \begin{align}
 \frac{\dot{f}}{f} = - 2 c H \left( 1 - \frac{\vep_H + \eta_H}{3}
 - \frac{c + 2}{3} I \right) + \calO (\vep I \,, \vep^2 \,, I^2)\,.
 \end{align}
%%%%%%%%%%%%%%%%%%%%%%%%%%%%%%%%%%%%%%%%%%%%%%%%%%%%%%%%%%%%%%%%%%%%%%%%%%
\subsection{Comparison with Watanabe, Kanno and Soda (2010)}
\label{app:soda}
By taking the logarithmic derivative of $I$ in equation~(\ref{def:I}) with
respect to $t$,
\begin{align}
 \frac{\dot{I}}{I}
 &= - \frac{\dot{\vep}_H}{\vep_H} - \frac{\dot{U}}{U}
 - 2 \frac{\dot{f}}{f} - 4 (H + \dot{\beta}) \,,
\end{align}
one obtains
 \begin{align}
 \frac{\dot{I}}{I}
 &= - 2 \frac{\dot{f}}{f} - 2 H (2 + \eta_H) + \calO (I)\,,
 \end{align}
 where we have assumed $U_\phi^2/(2 U^2) \approx \vep_H$.

Now, if we ignore $\dot{I}/I$, we have
\begin{align}
 \frac{\dot{f}}{f}
 = - H (2 + \eta_H) + \calO (I)\,.
\end{align}
This equation coincides exactly with the formula derived in
ref.~\cite{Watanabe:2010fh}. (Note that their definition of the symbol
$\vep_H$ is different from ours.) However, $\dot{I}/I$ gives rise to
terms of order $\vep_H$, which cannot be ignored, and thus ignoring it
leads to an erroneous result.
%%%%%%%%%%%%%%%%%%%%%%%%%%%%%%%%%%%%%%%%%%%%%%%%%%%%%%%%%%%%%%%%%%%%%%%%%%


\begin{thebibliography}{10}

\bibitem{Mukhanov:1990me}
V.~F. Mukhanov, H.~Feldman, and R.~H. Brandenberger, {\it {Theory of
  cosmological perturbations}},  {\em Phys.Rept.} {\bf 215} (1992) 203--333.

\bibitem{Mukhanov:1981xt}
V.~F. Mukhanov and G.~V. Chibisov, {\it {Quantum Fluctuation and Nonsingular
  Universe}},  {\em JETP Lett.} {\bf 33} (1981) 532--535.

\bibitem{Hinshaw:2012aka}
{\bf WMAP} Collaboration, G.~Hinshaw et~al., {\it {Nine-Year Wilkinson
  Microwave Anisotropy Probe (WMAP) Observations: Cosmological Parameter
  Results}},  {\em Astrophys.J.Suppl.} {\bf 208} (2013) 19,
  [\href{http://xxx.lanl.gov/abs/1212.5226}{\texttt{ arXiv:1212.5226}}].

\bibitem{Ade:2013zuv}
{\bf Planck Collaboration} Collaboration, P.~Ade et~al., {\it {Planck 2013
  results. XVI. Cosmological parameters}},  {\em Astron.Astrophys.} (2014)
  [\href{http://xxx.lanl.gov/abs/1303.5076}{\texttt{ arXiv:1303.5076}}].

\bibitem{Bennett:2012zja}
{\bf WMAP} Collaboration, C.~Bennett et~al., {\it {Nine-Year Wilkinson
  Microwave Anisotropy Probe (WMAP) Observations: Final Maps and Results}},
  {\em Astrophys.J.Suppl.} {\bf 208} (2013) 20,
  [\href{http://xxx.lanl.gov/abs/1212.5225}{\texttt{ arXiv:1212.5225}}].

\bibitem{Ade:2013ydc}
{\bf Planck Collaboration} Collaboration, P.~Ade et~al., {\it {Planck 2013
  Results. XXIV. Constraints on primordial non-Gaussianity}},  {\em
  Astron.Astrophys.} {\bf 571} (2014) A24,
  [\href{http://xxx.lanl.gov/abs/1303.5084}{\texttt{ arXiv:1303.5084}}].

\bibitem{Dimastrogiovanni:2010sm}
E.~Dimastrogiovanni, N.~Bartolo, S.~Matarrese, and A.~Riotto, {\it
  {Non-Gaussianity and Statistical Anisotropy from Vector Field Populated
  Inflationary Models}},  {\em Adv.Astron.} {\bf 2010} (2010) 752670,
  [\href{http://xxx.lanl.gov/abs/1001.4049}{\texttt{ arXiv:1001.4049}}].

\bibitem{Maleknejad:2012fw}
A.~Maleknejad, M.~Sheikh-Jabbari, and J.~Soda, {\it {Gauge Fields and
  Inflation}},  {\em Phys.Rept.} {\bf 528} (2013) 161--261,
  [\href{http://xxx.lanl.gov/abs/1212.2921}{\texttt{ arXiv:1212.2921}}].

\bibitem{Soda:2012zm}
J.~Soda, {\it {Statistical Anisotropy from Anisotropic Inflation}},  {\em
  Class.Quant.Grav.} {\bf 29} (2012) 083001,
  [\href{http://xxx.lanl.gov/abs/1201.6434}{\texttt{ arXiv:1201.6434}}].

\bibitem{Ackerman:2007nb}
L.~Ackerman, S.~M. Carroll, and M.~B. Wise, {\it {Imprints of a Primordial
  Preferred Direction on the Microwave Background}},  {\em Phys.Rev.} {\bf D75}
  (2007) 083502, [\href{http://xxx.lanl.gov/abs/astro-ph/0701357}{\texttt{
  astro-ph/0701357}}].

\bibitem{Watanabe:2010fh}
M.-a. Watanabe, S.~Kanno, and J.~Soda, {\it {The Nature of Primordial
  Fluctuations from Anisotropic Inflation}},  {\em Prog.Theor.Phys.} {\bf 123}
  (2010) 1041--1068, [\href{http://xxx.lanl.gov/abs/1003.0056}{\texttt{
  arXiv:1003.0056}}].

\bibitem{Barnaby:2012tk}
N.~Barnaby, R.~Namba, and M.~Peloso, {\it {Observable non-gaussianity from
  gauge field production in slow roll inflation, and a challenging connection
  with magnetogenesis}},  {\em Phys.Rev.} {\bf D85} (2012) 123523,
  [\href{http://xxx.lanl.gov/abs/1202.1469}{\texttt{ arXiv:1202.1469}}].

\bibitem{Bartolo:2012sd}
N.~Bartolo, S.~Matarrese, M.~Peloso, and A.~Ricciardone, {\it {The anisotropic
  power spectrum and bispectrum in the $f(\phi) F^2$ mechanism}},  {\em
  Phys.Rev.} {\bf D87} (2013) 023504,
  [\href{http://xxx.lanl.gov/abs/1210.3257}{\texttt{ arXiv:1210.3257}}].

\bibitem{Shiraishi:2013vja}
M.~Shiraishi, E.~Komatsu, M.~Peloso, and N.~Barnaby, {\it {Signatures of
  anisotropic sources in the squeezed-limit bispectrum of the cosmic microwave
  background}},  {\em JCAP} {\bf 1305} (2013) 002,
  [\href{http://xxx.lanl.gov/abs/1302.3056}{\texttt{ arXiv:1302.3056}}].

\bibitem{Shiraishi:2013oqa}
M.~Shiraishi, E.~Komatsu, and M.~Peloso, {\it {Signatures of anisotropic
  sources in the trispectrum of the cosmic microwave background}},  {\em JCAP}
  {\bf 1404} (2014) 027, [\href{http://xxx.lanl.gov/abs/1312.5221}{\texttt{
  arXiv:1312.5221}}].

\bibitem{Emami:2013bk}
R.~Emami and H.~Firouzjahi, {\it {Curvature Perturbations in Anisotropic
  Inflation with Symmetry Breaking}},  {\em JCAP} {\bf 1310} (2013) 041,
  [\href{http://xxx.lanl.gov/abs/1301.1219}{\texttt{ arXiv:1301.1219}}].

\bibitem{Abolhasani:2013zya}
A.~A. Abolhasani, R.~Emami, J.~T. Firouzjaee, and H.~Firouzjahi, {\it {$\delta
  N$ formalism in anisotropic inflation and large anisotropic bispectrum and
  trispectrum}},  {\em JCAP} {\bf 1308} (2013) 016,
  [\href{http://xxx.lanl.gov/abs/1302.6986}{\texttt{ arXiv:1302.6986}}].

\bibitem{Emami:2014tpa}
R.~Emami, H.~Firouzjahi, and M.~Zarei, {\it {Anisotropic Inflation with the
  non-Vacuum Initial State}},  {\em Phys.Rev.} {\bf D90} (2014) 023504,
  [\href{http://xxx.lanl.gov/abs/1401.4406}{\texttt{ arXiv:1401.4406}}].

\bibitem{Dey:2011mj}
A.~Dey and S.~Paban, {\it {Non-Gaussianities in the Cosmological Perturbation
  Spectrum due to Primordial Anisotropy}},  {\em JCAP} {\bf 1204} (2012) 039,
  [\href{http://xxx.lanl.gov/abs/1106.5840}{\texttt{ arXiv:1106.5840}}].

\bibitem{Dey:2012qp}
A.~Dey, E.~Kovetz, and S.~Paban, {\it {Non-Gaussianities in the Cosmological
  Perturbation Spectrum due to Primordial Anisotropy II}},  {\em JCAP} {\bf
  1210} (2012) 055, [\href{http://xxx.lanl.gov/abs/1205.2758}{\texttt{
  arXiv:1205.2758}}].

\bibitem{Dey:2013tfa}
A.~Dey, E.~D. Kovetz, and S.~Paban, {\it {Power Spectrum and Non-Gaussianities
  in Anisotropic Inflation}},  {\em JCAP} {\bf 1406} (2014) 025,
  [\href{http://xxx.lanl.gov/abs/1311.5606}{\texttt{ arXiv:1311.5606}}].

\bibitem{Endlich:2012pz}
S.~Endlich, A.~Nicolis, and J.~Wang, {\it {Solid Inflation}},  {\em JCAP} {\bf
  1310} (2013) 011, [\href{http://xxx.lanl.gov/abs/1210.0569}{\texttt{
  arXiv:1210.0569}}].

\bibitem{Bartolo:2013msa}
N.~Bartolo, S.~Matarrese, M.~Peloso, and A.~Ricciardone, {\it {Anisotropy in
  solid inflation}},  {\em JCAP} {\bf 1308} (2013) 022,
  [\href{http://xxx.lanl.gov/abs/1306.4160}{\texttt{ arXiv:1306.4160}}].

\bibitem{Bartolo:2014xfa}
N.~Bartolo, M.~Peloso, A.~Ricciardone, and C.~Unal, {\it {The expected
  anisotropy in solid inflation}},  {\em JCAP} {\bf 1411} (2014) 009,
  [\href{http://xxx.lanl.gov/abs/1407.8053}{\texttt{ arXiv:1407.8053}}].

\bibitem{Akhshik:2014gja}
M.~Akhshik, R.~Emami, H.~Firouzjahi, and Y.~Wang, {\it {Statistical
  Anisotropies in Gravitational Waves in Solid Inflation}},  {\em JCAP} {\bf
  1409} (2014) 012, [\href{http://xxx.lanl.gov/abs/1405.4179}{\texttt{
  arXiv:1405.4179}}].

\bibitem{Watanabe:2009ct}
M.-a. Watanabe, S.~Kanno, and J.~Soda, {\it {Inflationary Universe with
  Anisotropic Hair}},  {\em Phys.Rev.Lett.} {\bf 102} (2009) 191302,
  [\href{http://xxx.lanl.gov/abs/0902.2833}{\texttt{ arXiv:0902.2833}}].

\bibitem{Dimopoulos:2010xq}
  J.~M.~Wagstaff and K.~Dimopoulos, {\it {Particle Production of
 Vector Fields: Scale Invariance is Attractive}}, 
 {\em Phys.Rev.} {\bf D83} (2011) 023523,
  [\href{http://xxx.lanl.gov/abs/1011.2517}{\texttt{ arXiv:1011.2517}}].

\bibitem{Kim:2013gka}
J.~Kim and E.~Komatsu, {\it {Limits on anisotropic inflation from the Planck
  data}},  {\em Phys.Rev.} {\bf D88} (2013) 101301,
  [\href{http://xxx.lanl.gov/abs/1310.1605}{\texttt{ arXiv:1310.1605}}].

\bibitem{Creminelli:2012ed}
P.~Creminelli, J.~Norena, and M.~Simonovic, {\it {Conformal consistency
  relations for single-field inflation}},  {\em JCAP} {\bf 1207} (2012) 052,
  [\href{http://xxx.lanl.gov/abs/1203.4595}{\texttt{ arXiv:1203.4595}}].

\bibitem{Maleknejad:2012as}
A.~Maleknejad and M.~Sheikh-Jabbari, {\it {Revisiting Cosmic No-Hair Theorem
  for Inflationary Settings}},  {\em Phys.Rev.} {\bf D85} (2012) 123508,
  [\href{http://xxx.lanl.gov/abs/1203.0219}{\texttt{ arXiv:1203.0219}}].

\bibitem{Cheung:2007st}
C.~Cheung, P.~Creminelli, A.~L. Fitzpatrick, J.~Kaplan, and L.~Senatore, {\it
  {The Effective Field Theory of Inflation}},  {\em JHEP} {\bf 0803} (2008)
  014, [\href{http://xxx.lanl.gov/abs/0709.0293}{\texttt{ arXiv:0709.0293}}].

\bibitem{Cannone:2014uqa}
D.~Cannone, G.~Tasinato, and D.~Wands, {\it {Generalised tensor fluctuations
  and inflation}},  \href{http://xxx.lanl.gov/abs/1409.6568}{\texttt{
  arXiv:1409.6568}}.
\end{thebibliography}
\end{document}